\documentclass[12pt,14paper]{article}
\usepackage{amsmath,amsfonts}
\usepackage{epsfig}
\usepackage{array}
\usepackage{float}
\usepackage{lscape,graphicx} 
\usepackage{dcolumn}
\usepackage{graphics}
\usepackage{amssymb}
\usepackage{color}
\usepackage{multirow}
\usepackage{advdate}
\usepackage{datenumber}
\usepackage{makecell} 
\usepackage{colortbl}
\usepackage{relsize}
\usepackage{placeins}
\usepackage{slashed}
\usepackage{ulem}
\usepackage[dvipsnames]{xcolor}

\definecolor{LightCyan}{rgb}{0.88,1,1}
\definecolor{Blond}{rgb}{0.98, 0.94, 0.75}
\definecolor{beige}{rgb}{0.96, 0.96, 0.86}
\definecolor{champagne}{rgb}{0.97, 0.91, 0.81}

\newcommand{\bea}{\begin{equation}\begin{array}{c}}
\newcommand{\eea}{\end{array}\end{equation}}
\newcommand{\ea}{\end{array}}

\newcommand{\beq}{\begin{equation}}
\newcommand{\eeq}{\end{equation}}
\newcommand{\bad}{\begin{array}{ccc}}

\newcommand{\ba}{\begin{array}{c}}

\newcommand{\rhomean}{\langle\rho\rangle}
\newcommand{\ZAmean}{\langle Z/A\rangle}

\newcommand{\dedxmeanminus}{\left\langle-dE/dx\right\rangle}

\newcommand{\ssp}{\hspace{.2mm}} 
\newcommand{\Acc}{\mathcal{A}}

\newcommand{\dmnc}{\mbox{$\Delta m_\mathrm{nc}$}}

\newcommand{\half}{\frac{1}{2}}

\newcommand{\Hpm}{{H^\pm}}

\newcommand{\Imean}{\langle I\rangle}

\newcommand{\Estop}{{\epsilon_\mathrm{stop}}}
\newcommand{\Erec}{{\epsilon_\mathrm{rec}}}

\newcommand{\Etiming}{{\epsilon_\mathrm{T}}}

\begin{document}

\title{
{\bf Probing the scotogenic FIMP at the LHC}}

\author{Andre~G.~Hessler$^1\footnote{andre.hessler@tum.de}$, Alejandro~Ibarra$^1\footnote{ibarra@tum.de}$, Emiliano~Molinaro$^{2}\footnote{molinaro@cp3-origins.net}$, and Stefan~Vogl$^{3,4}\footnote{stefan.vogl@mpi-hd.mpg.de}$\\[7mm]
\it \normalsize $^1$Physik-Department T30d, Technische Universit\"at M\"unchen,\\ 
\it \normalsize James-Franck-Stra{\ss}e, 85748 Garching, Germany\\[2mm]
\it \normalsize $^2$CP$^3$-Origins and University of Southern Denmark, \\ 
\it \normalsize Campusvej 55, DK-5230 Odense M, Denmark\\[2mm]
\it \normalsize  $^3$Max Planck Institute for Nuclear Physics,\\ 
\it \normalsize Saupfercheckweg 1, 69117 Heidelberg, Germany\\[2mm]
\it \normalsize  $^4$Institute for Nuclear Physics, Karlsruhe Institute of Technology,\\ 
\it \normalsize Hermann-von-Helmholtz-Platz 1, 76344 Eggenstein-Leopoldshafen, Germany
}
\date{}
\maketitle
\vspace*{-12cm}
\begin{flushright}
\texttt{\footnotesize TUM-HEP-1068/16}\\[-1mm] 
\texttt{\footnotesize CP3-Origins-2016-051 DNRF90}\\[-1mm] 
\end{flushright}
\vspace*{9.5cm}
\thispagestyle{empty}
\begin{abstract}
We analyse the signatures at the Large Hadron Collider (LHC) of the scotogenic model, when the  lightest $Z_2$-odd particle is  a singlet fermion and a feebly interacting massive particle (FIMP). We further assume that the singlet fermion constitutes the dark matter and that it is produced  in the early Universe via the freeze-in mechanism. The small couplings required to reproduce the observed dark matter abundance translate into decay-lengths for the next-to-lightest $Z_2$-odd particle which can be macroscopic, potentially leading to spectacular signatures at the LHC. We characterize the possible signals of the model according to the spectrum of the $Z_2$-odd particles and we derive, for each of the cases, bounds on the parameters of the model from current searches.
\end{abstract}

\section{Introduction}

Understanding the nature of the dark matter of the Universe stands among the most pressing problems in  Astroparticle Physics. A plausible hypothesis is that the dark matter is constituted by a population of new particles not contained in the Standard Model (for reviews, see ~\cite{Bertone:2010zza,Bergstrom:2000pn,Bertone:2004pz}). Testing this hypothesis is, however, impeded by the vast number of dark matter candidates proposed in the literature, with very disparate characteristics and, correspondingly, with very different experimental signatures~\cite{Baer:2014eja}. Over many years, the most popular and most studied dark matter candidate has been the Weakly Interacting Massive Particle (WIMP) \cite{Gershtein:1966gg}. In this framework, the dark matter is assumed to be in thermal and chemical equilibrium with the plasma of Standard Model particles, and to become decoupled when the expansion rate becomes faster than the annihilation rate. If the strength of the WIMP interactions with the Standard Model particles is comparable to that of the weak interactions, the WIMP relic abundance is predicted to be in the ballpark of the observed dark matter abundance $\Omega h^2\simeq 0.12$~\cite{Ade:2013zuv}. The simplicity and economy of this framework has triggered enormous experimental efforts to detect non-gravitational WIMP signals. Unfortunately, no unambiguous signal has been found yet. 

The lack of evidence for WIMP dark matter has prompted interest in other dark matter candidates and  their identification, which usually requires search strategies very different to those for WIMPs. In this paper we concentrate on FIMPs as dark matter candidates~\cite{Hall:2009bx}. FIMPs are characterized by a small interaction rate with the Standard Model particles, so that they never reach thermal equilibrium with the SM plasma throughout the whole cosmological history. However, they can be produced in decays and annihilations of Standard Model particles in the thermal bath, leading to a relic abundance which is essentially dictated by the time at which the expansion rate exceeded the production rate.

More concretely, we focus on a specific FIMP realization which may be linked to the mechanism of neutrino mass generation. We consider the scotogenic model~\cite{Ma:2006km}, where the Standard Model particle content is extended by three fermion singlets and one scalar doublet. The model further assumes that the new particles are odd under a discrete $Z_2$ symmetry, assumed to be exactly conserved in the electroweak vacuum, while all the Standard Model particles are even. The lightest particle of the $Z_2$-odd sector is absolutely stable and therefore constitutes a dark matter candidate. Several works have analysed the phenomenology of the scotogenic model when the dark matter candidate is a WIMP, which can be identified either with the lightest $Z_2$-odd scalar~\cite{Ma:2006fn} or with the lightest  $Z_2$-odd fermion~\cite{Kubo:2006yx}. In this work, in contrast, we will consider the scenario where the dark matter is a FIMP~\cite{Molinaro:2014lfa}. Due to the tiny strength of their interactions with ordinary matter, no observable signal is expected in direct and indirect search experiments. We will show in this paper that very distinctive signals may arise at the Large Hadron Collider, thus offering a promising avenue to test this model. 

The paper is organized as follows. In Section \ref{sec:scot} we review the main characteristics of the FIMP dark matter in the scotogenic model, in Section \ref{sec:Collider-pheno} we scrutinize the collider phenomenology of the model for three representative choices of the mass spectrum of the $Z_2$-odd particles and in Section \ref{sec:Conclusions} we present our conclusions.

\section{FIMP dark matter in the scotogenic model}
\label{sec:scot}

In the scotogenic model  \cite{Ma:2006km}, the particle content of the Standard Model (SM) is extended with one additional scalar doublet $H_2\equiv (H^+,H_2^0)$ and three fermionic singlets $N_j$ ($j=1,2,3$). The model also postulates that the electroweak vacuum is invariant under a discrete $Z_2$ symmetry, under which all SM fields are even, whereas $N_j$ and $H_2$ are odd. The Lagrangian of the model is given by
\begin{align}
\mathcal{L} =  \mathcal{L}_{\rm SM}\,+ \mathcal{L}_{H_2}\,+ \mathcal{L}_{N}\,+ \mathcal{L}_{\rm int}\;,
\end{align}
where $\mathcal{L}_{\rm SM}$ denotes the SM Lagrangian, which includes the potential for the SM Higgs doublet $H_1$,
\begin{align}
{\cal L}_{\rm SM}\supset -\,\mu_{1}^{2}\,(H_{1}^{\dagger}\,H_{1})\,+\,\lambda_{1}\,(H_{1}^{\dagger}\,H_{1})^{2}\;,
\end{align}
$\mathcal{L}_{H_2}$ and $\mathcal{L}_{N}$ are, respectively, the terms in the Lagrangian involving only the fields $H_2$ and~$N_j$,
\begin{align}
\mathcal{L}_{H_2} & =  \left(D_\mu H_2\right)^\dagger(D^\mu H_2)+\,\mu_{2}^{2}\,(H_{2}^{\dagger}\,H_{2})\,+\,\lambda_{2}\,(H_{2}^{\dagger}\,H_{2})^{2}\;, \\
\mathcal{L}_{N} &= \frac{i}{2}\,\overline{N}_j\partial_\mu\gamma^\mu N_j-\half\,M_{j}\,\overline{N^c_j} \,N_{j}+{\rm h.c.}\;,
\end{align}
and $\mathcal{L}_{\rm int}$ contains the interaction terms of the $Z_2$-odd fields with the Standard Model fields,
\begin{align}
\mathcal{L}_{\rm int}=&\,\lambda_{3}\,(H_{1}^{\dagger}\,H_{1})\,(H_{2}^{\dagger}\,H_{2})
	\,+\,\lambda_{4}\,(H_{1}^{\dagger}\,H_{2})\,(H_{2}^{\dagger}\,H_{1})\,+\, \frac{\lambda_{5}}{2}\,\left[(H_{1}^{\dagger}\,H_{2})^{2}\,+\,{\rm h.c.}\right] \nonumber \\
&+\Big[Y^{\nu}_{\alpha i}\,(\overline{\nu}_{\alpha L}\,H_{2}^{0}\,-\,\overline{\ell}_{\alpha L}\,H^{+})\,N_{i}	 \,+\,{\rm h.c.}\Big] \;.
\label{eq:L_int}
\end{align}
The parameters of the scalar potential are chosen such that $\langle H_{1}\rangle =(0, v/\sqrt{2})$, with $v\simeq 246$ GeV, and  $\langle H_{2}\rangle =0$, hence the minimum of the potential breaks the electroweak symmetry while preserving the  $Z_2$ symmetry. 

The multiplet $H_1$ contains only one physical scalar state, the SM Higgs boson $h$, with $m_h=125$ GeV  \cite{Aad:2012tfa,Chatrchyan:2012ufa}. On the other hand, the $Z_2$-odd scalar sector contains one CP-even ($H^0$), one CP-odd ($A^0$) and two charged ($H^\pm$) scalar fields, with masses 
\begin{align}\label{eq:scalar_masses}
	\begin{split}
	     m_{H^{0}}^{2} & = \; \mu_{2}^{2} \,+\,v^{2}\,\left(\lambda_{3}+\lambda_{4}+\lambda_{5}\right)/2\,, \\
	    m_{A^{0}}^{2}  & = \; \mu_{2}^{2} \,+\,v^{2}\,\left(\lambda_{3}+\lambda_{4}-\lambda_{5}\right)/2\,, \\
	   m_{H^{\pm}}^{2} & = \; \mu_{2}^{2}\,+\,v^{2}\,\lambda_{3}/2\,,
	   \end{split}
\end{align}
 where $\mu_2$ and $\lambda_{3,4,5}$ are couplings in the scalar potential.

The requirement of stability of the scalar potential  further constrains the quartic couplings $\lambda_k$ (see, e.g., \cite{Hambye:2009pw}), which must satisfy the conditions 
\begin{align}\label{eq:stability}
	\begin{split}
           &		\lambda_{1,2} \,>\, 0 \,,\\
           &		\lambda_3 \,>\, -\sqrt{\lambda_1\,\lambda_2}\,,\\
           &		\lambda_3\,+\lambda_4\,\pm \left|\lambda_5\right| \,>\,-\sqrt{\lambda_1\,\lambda_2}\,.
           \end{split}
\end{align}
Additional relations involving different combinations of $\lambda_k$ arise from perturbative partial-wave unitarity of tree-level scattering diagrams \cite{Ginzburg:2003fe}.

The model violates total lepton number and, in general, all three family lepton numbers. However, due to the assignment of the $Z_2$ charges, lepton and flavour violating processes involving only Standard Model particles in the initial and final states only arise at the one loop level, and thus with a suppressed rate. In particular, the model predicts non-vanishing neutrino masses. At lowest order in perturbation theory the neutrino mass term reads~\cite{Ma:2006km, Merle:2015ica}:
\begin{align}
\label{eq:masses}
	\left(\mathcal{M}_{\nu}\right)_{\alpha \beta} &~= \sum_{k}\,\frac{Y^{\nu}_{\alpha k}\,Y^{\nu}_{\beta k}}{32\,\pi^{2}}\,M_{k}\,
	\left[\frac{m_{H^{0}}^{2}}{m_{H^{0}}^{2}-M_{k}^{2}}\log\left(\frac{m_{H^{0}}^{2}}{M_{k}^{2}}\right)\,-\,
	\frac{m_{A^{0}}^{2}}{m_{A^{0}}^{2}-M_{k}^{2}}\log\left(\frac{m_{A^{0}}^{2}}{M_{k}^{2}}\right)\right]\;.
\end{align}

Small neutrino masses can be generated by {\it i)} postulating small Yukawa couplings, {\it ii)} postulating that the $Z_2$-odd particles in the loop are heavy, {\it iii)} postulating that the quartic coupling $\lambda_5$ is very small. Of special interest is the scenario where the new particles responsible for neutrino masses have masses below the TeV scale and sizeable couplings with the Standard Model particles. If this is the case, the new particles could produce observable signals in experiments at the energy and the intensity frontiers, thus opening the exciting possibility of testing the model of neutrino mass generation. 
For $\lambda_5 \ll 1$ this  expression simplifies and the mass matrix can be written as
\begin{align}
	\left(\mathcal{M}_{\nu}\right)_{\alpha \beta} \simeq  \frac{\lambda_{5}\,v^{2}}{32\,\pi^{2}}\,\sum_{k}\,Y^{\nu}_{\alpha k}\,Y^{\nu}_{\beta k}\,\frac{M_{k}}{m_{0}^{2}-M_{k}^{2}}\,
	\left[1-\frac{M_{k}^{2}}{m_{0}^{2}-M_{k}^{2}}\,\log\left(\frac{m_{0}^{2}}{M_{k}^{2}}\right)\right]\,,
\label{eq:mn}
\end{align}
where  $m_{0}^{2}=\left(m_{H^{0}}^{2}+m_{A^{0}}^{2}\right)/2$. This assumption implies in particular, $m_{H^{0}}^{2}\simeq m_{A^{0}}^{2}$ and $m_{A_0}^2-m_{H^\pm}^2\simeq v^2 \lambda_4/2$, as follows from Eq.~(\ref{eq:scalar_masses}). 

The rates for the charged lepton flavour violating processes also vanish at tree level, but are predicted to arise at the one loop level \cite{Kubo:2006yx,Toma:2013zsa}.  The strongest limits on the model parameters follow from the current upper bound on the $\mu\to e\,\gamma$ branching ratio, $\text{BR}(\mu\to e\,\gamma)<4.2\times 10^{-13}$ at 90\% CL  \cite{TheMEG:2016wtm}. In the scotogenic model, the branching ratio reads
\begin{align}\label{eq:brmueg}
	\begin{split}
	\text{BR}(\mu\to e\,\gamma) & = \; \frac{3\,\alpha_{\rm em}}{64\,\pi\, G^2_{F}\, m_{H{^\pm}}^4}\,\left|\sum_k Y^{\nu}_{\mu k}\, Y{^{\nu}_{e k}}^{*}\, F_{2}\left(\frac{M_{k}^{2}}{m_{H^{\pm}}^{2}}\right) \right|^{2}\\
	&\approx\; 10^{-15}\left(\frac{100\,\mbox{GeV}}{m_H^{\pm}}\right)^4\,\left|\frac{y_{k}}{10^{-2}}\right|^4\,\left(\frac{F_2(M_{k}^2/m_{H^{\pm}}^2)}{3\times10^{-3}}\right)^2\,,
	\end{split}
\end{align}
where $y_k\equiv (\sum_\alpha |Y_{\alpha k}|^2)^{1/2}$, $k=1,2,3$, and $F_2(x)$ is a monotonically decreasing loop function  which varies between $0.14$ and $3\times 10^{-5}$ for $x$ between $0.1$ and $10^{4}$.  

The conservation of the $Z_2$ symmetry ensures that the lightest $Z_2$-odd particle is absolutely stable, which then constitutes a dark matter candidate if it is electrically neutral. The dark matter candidates of the model are the CP-even and CP-odd neutral scalars, $H^0$ and $A^0$, and the lightest singlet fermion $N_1$. Here we focus on the latter candidate, the singlet fermion. The dark matter may be produced via the freeze-out mechanism, however this mechanism requires sizeable Yukawa couplings which generically lead to too large rates for the rare leptonic decays (see, however, \cite{Ibarra:2016dlb}). In order to suppress the rare decays we assume that $N_1$ interacts very weakly with the Standard Model particles. The dark matter population can then be generated from the freeze-in mechanism \cite{Hall:2009bx} and from the decay of the next-to-lightest  $Z_2$-odd particle, dubbed superWIMP mechanism \cite{Feng:2003xh}. The latter contribution, on the other hand, can always be neglected if the scalar particles have a mass $m_{H_2}\lesssim 500$ GeV and $M_1\lesssim 100$ MeV (see \cite{Molinaro:2014lfa} for a detailed discussion).

 In the following we will consider the region of the parameter space of the model
where the observed dark matter abundance, $\Omega_{\rm DM} h^2\approx 0.12$, is entirely generated by the freeze-in mechanism. The dark matter density is approximately given by  \cite{Molinaro:2014lfa}
\begin{align}
  \Omega_{N_{1}} h^{2}&\approx 0.12\left(\frac{M_{1}}{10\,\mbox{keV}}\right)\left(\frac{100\,\mbox{GeV}}{m_{H_2}}\right)\left(\frac{y_1}{2\times 10^{-9}}\right)^2,
\label{eq:rd}
\end{align}
provided $|m_{H^0/A^0}-m_{H^\pm}|\lesssim 100$ GeV, where $m_{H_2}$ is the overall mass of the $Z_2$-odd scalars.
Therefore, a ballpark estimate of the size of the Yukawa couplings leading to the observed dark matter abundance via freeze-in is
\begin{equation}
y_1  \approx 2\times10^{-9} \left(\frac{10~\mathrm{keV}}{M_1}\right)^{1/2} \left(\frac{m_{H_2}}{100~\mathrm{GeV}}\right)^{1/2}\,.\label{y1}
\end{equation}

The tiny Yukawa couplings required by the FIMP mechanism imply that the dark matter plays a subdominant role in the neutrino mass generation, which is then dominated by the heavier fermions $N_{2,3}$. The strength of their Yukawa interactions to the Standard Model is constrained from below by the requirement of correctly reproducing the measured neutrino parameters, and constrained from above by the experimental upper bound on ${\rm BR}(\mu\rightarrow e\gamma)$. Postulating masses for the new particles below the TeV scale and assuming $\lambda_5 < 0.1$, one obtains 
\begin{equation}
     10^{-5}	\; \lesssim \;  y_{2,3} \; \lesssim\;  10^{-2}\,.\label{boundy23}
\end{equation}

\begin{figure}[t]
\centering
\begin{tabular}{ccc}
\includegraphics[width=0.23\textwidth]{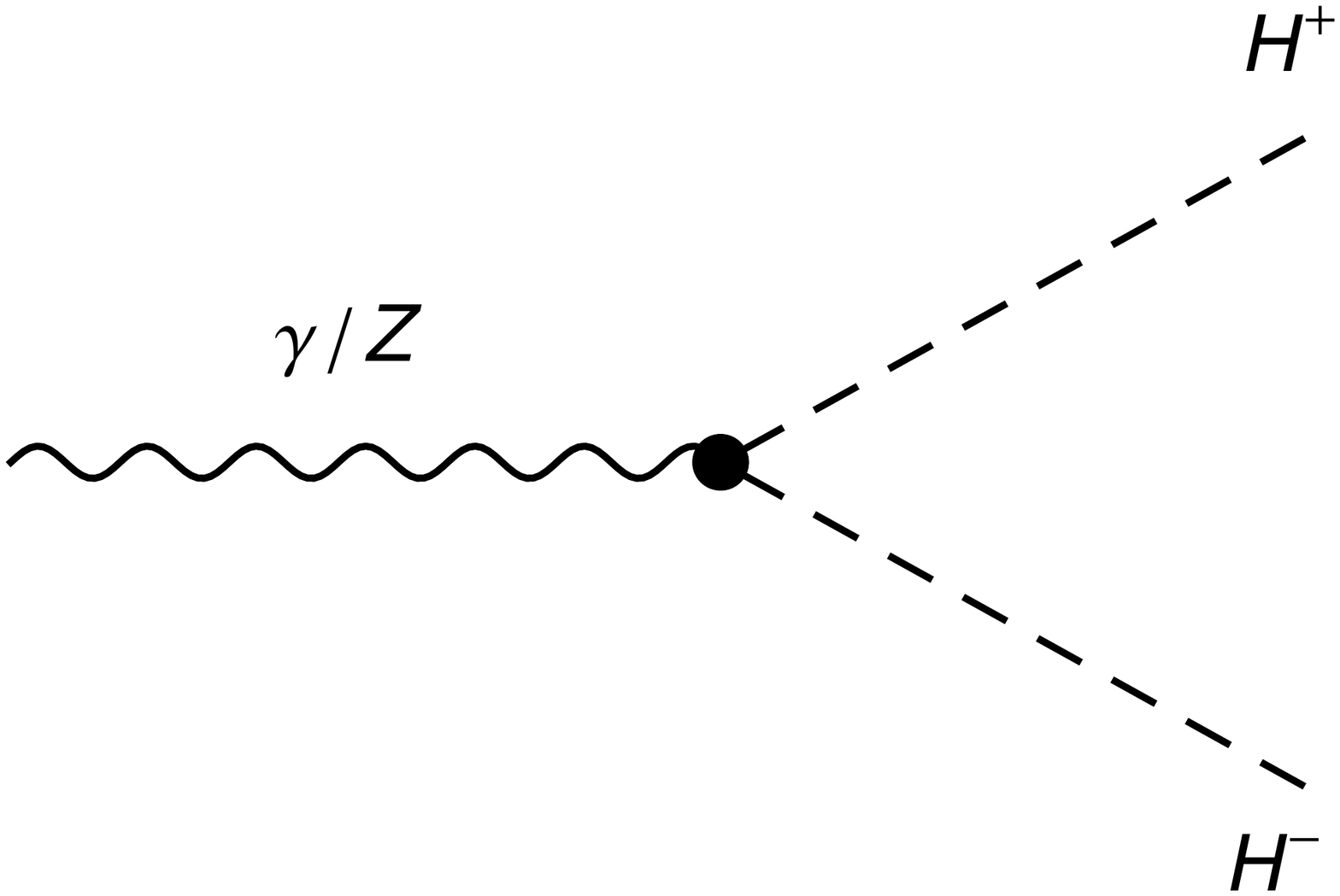} &
\includegraphics[width=0.23\textwidth]{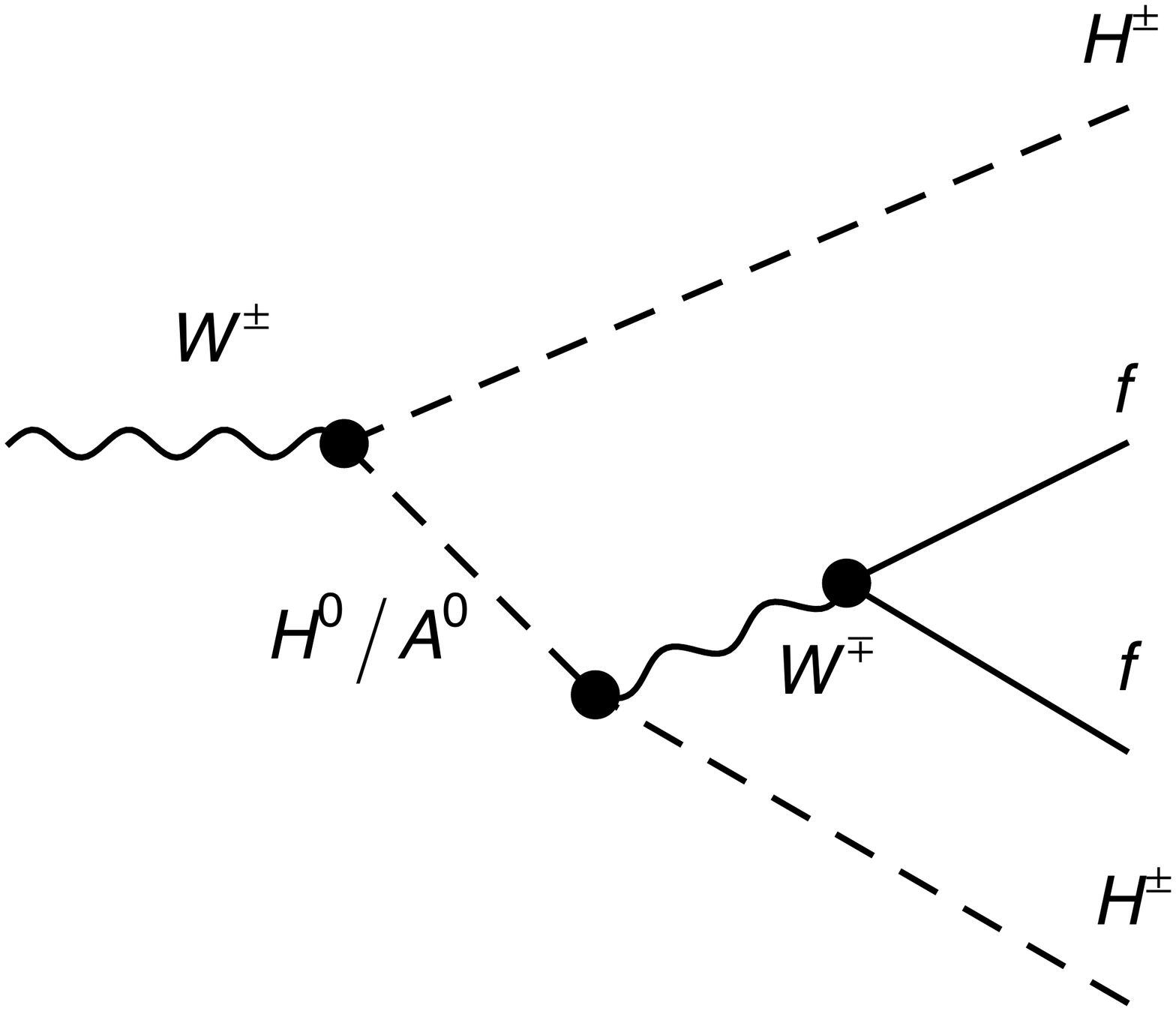} &
\includegraphics[width=0.23\textwidth]{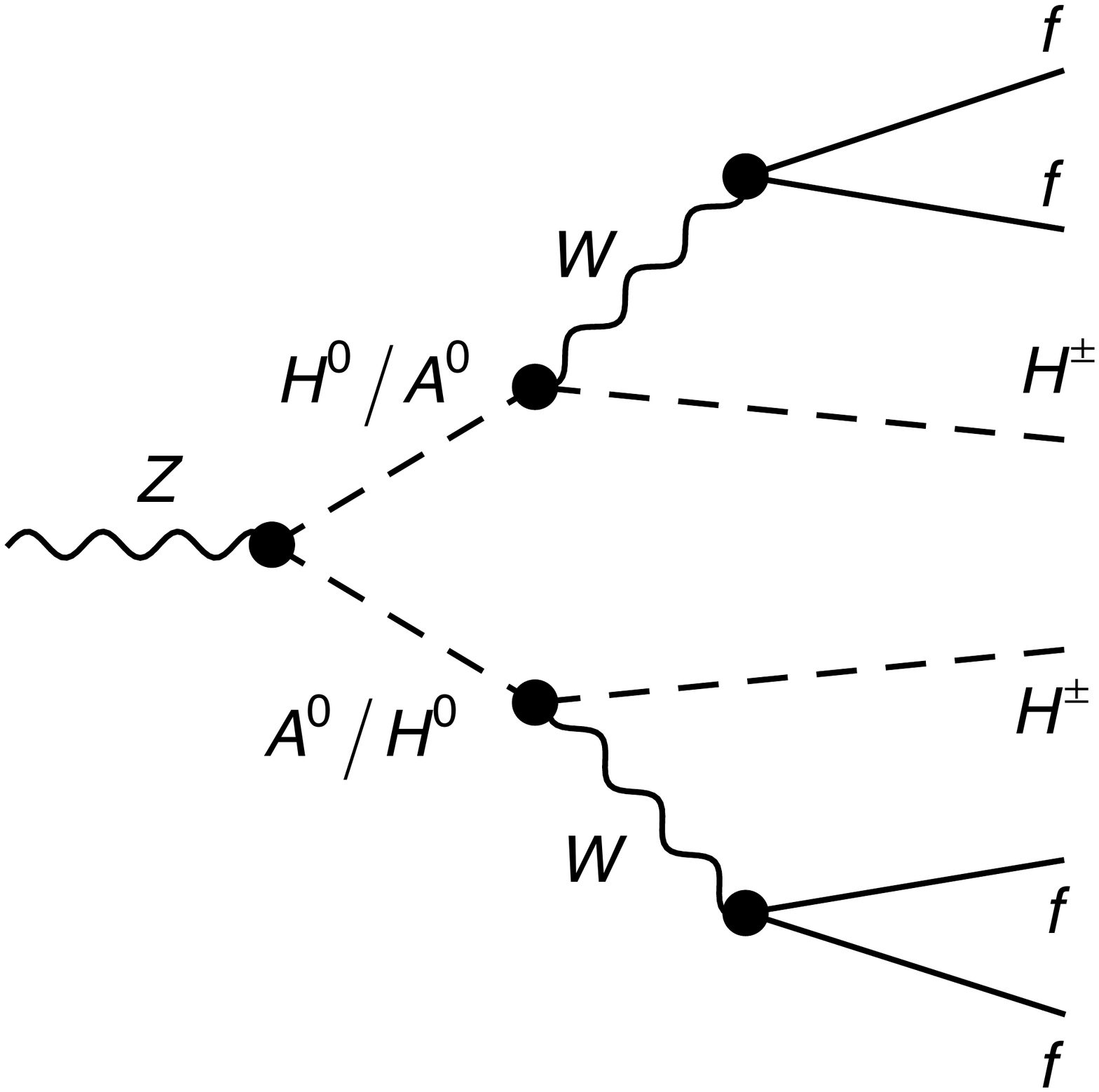}\\
\vspace{-2cm}\\
\multicolumn{3}{c}{
\includegraphics[width=0.23\textwidth]{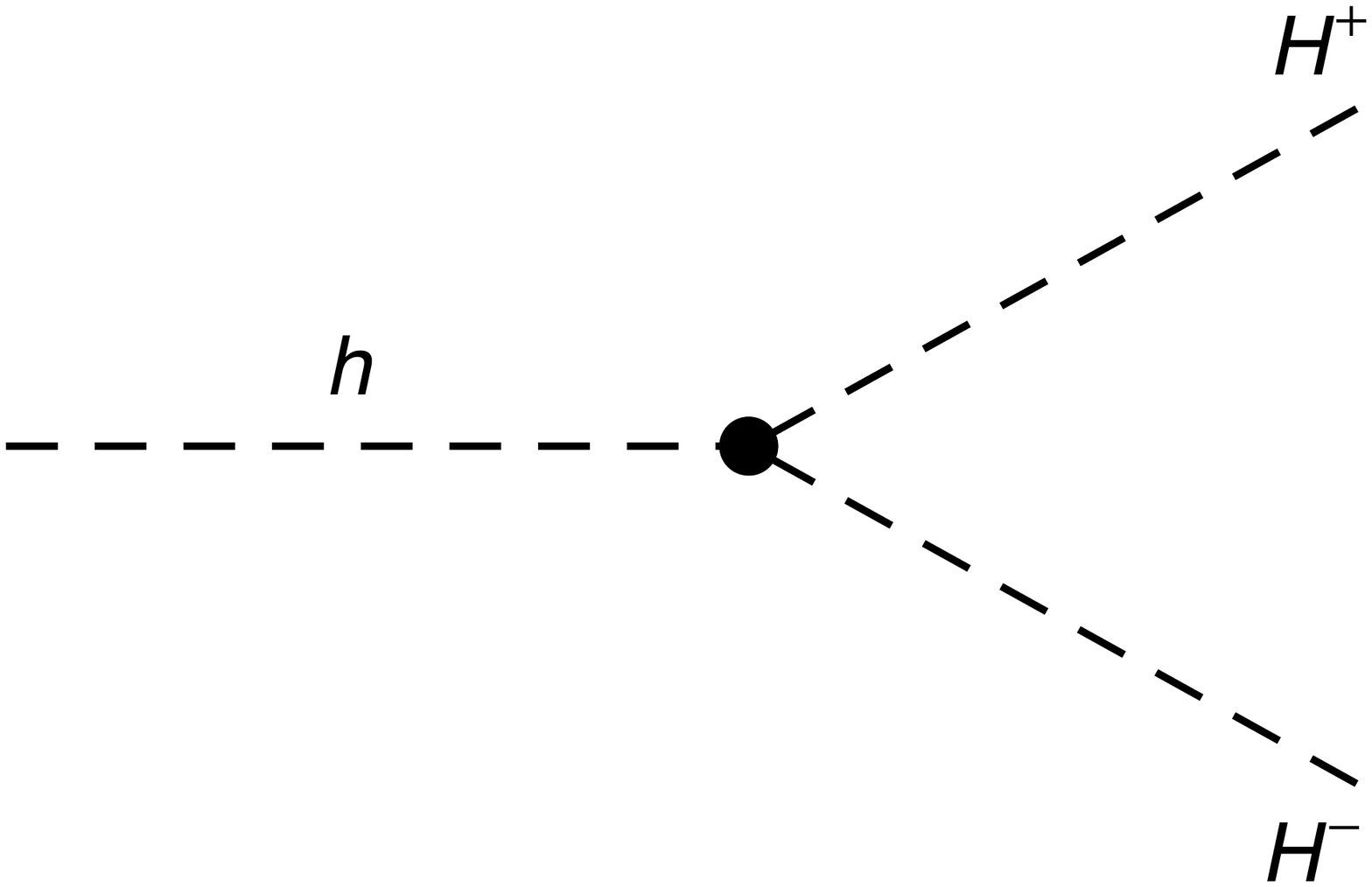}
\includegraphics[width=0.23\textwidth]{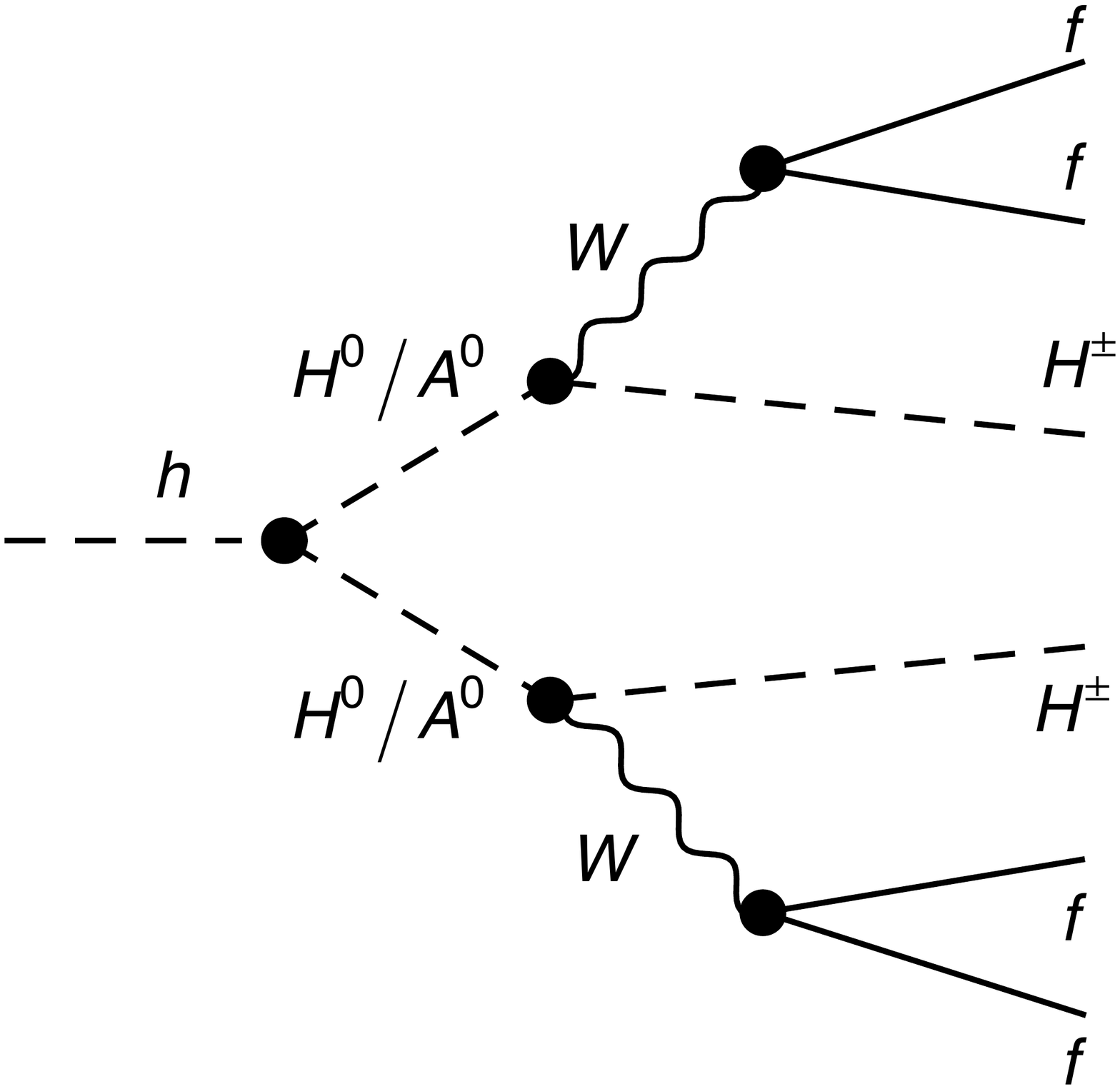}
}\\
\vspace{-1.5cm}
\end{tabular}
\caption{\small{
Drell-Yan (upper row) and gluon-gluon fusion (lower row) production channels of the $Z_2$-odd charged scalar.}}
\label{Fig:SpecAHcProdProcesses}
\end{figure}

\section{Collider phenomenology}
\label{sec:Collider-pheno}
\begin{figure}[t!]
\begin{center}
\includegraphics[width=0.7\textwidth]{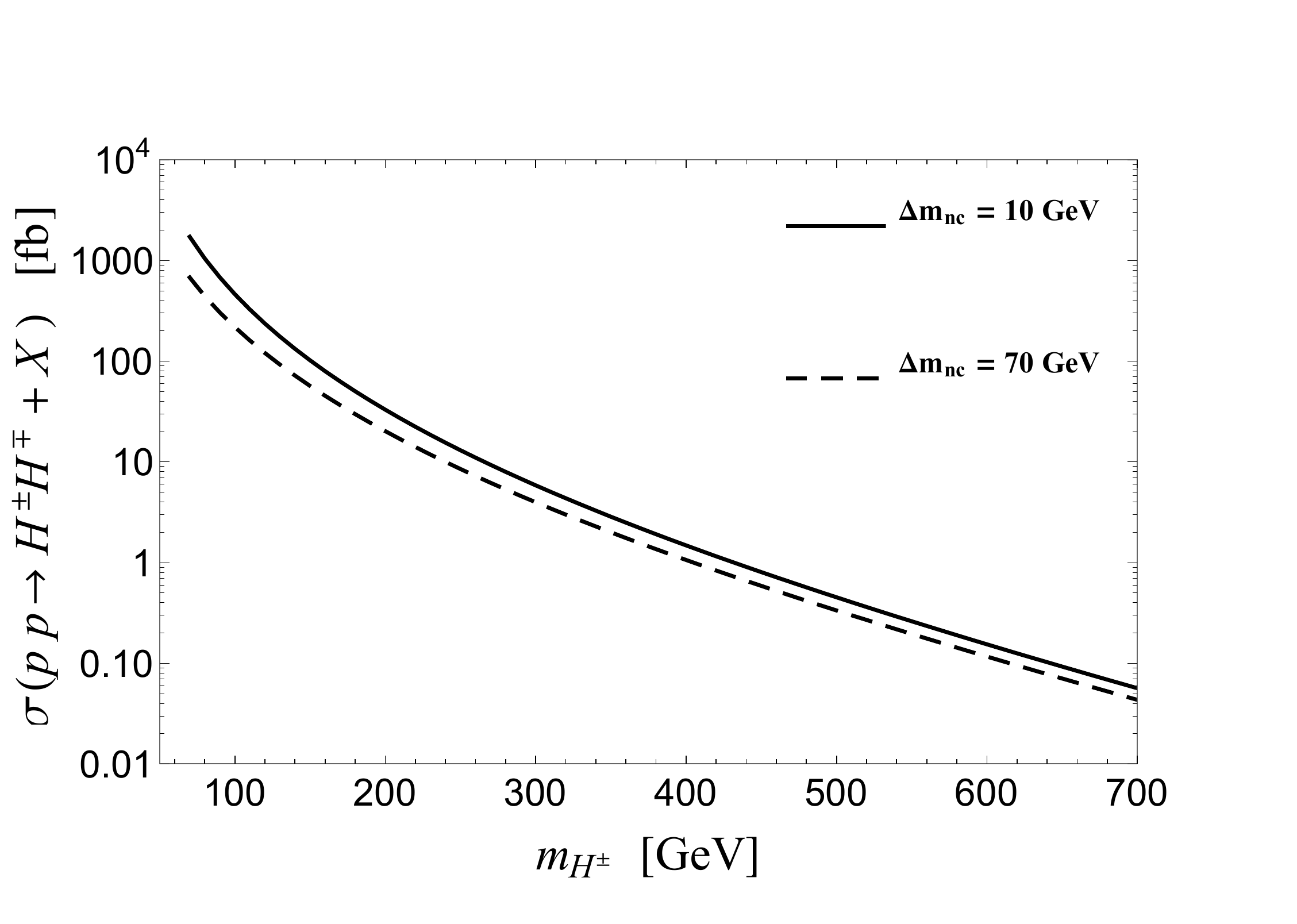} 
\caption{{\small  
Drell-Yan production cross-sections at $\sqrt{s}=8$ TeV  for $\dmnc=10$ GeV (black solid line) and $\dmnc=70$  GeV (black dashed line).
 }}
\label{Fig:HcProdCMSlimit}
\end{center}
\end{figure}

FIMPs, due to their tiny interactions with the Standard Model particles, are not directly produced at colliders. However, they are produced in the decays of the $Z_2$-odd scalars $H^0, \,A^0$ and $H^\pm$, which can be copiously produced at the LHC  via neutral and charged current Drell-Yan  (DY)  processes. \footnote{Under certain conditions gluon-gluon fusion (ggF) with an off-shell Higgs in the s-channel can also be relevant~\cite{Hessler:2014ssa}.} Here we assume for concreteness that $m_{H^\pm}< m_{H^0, A^0}$, corresponding to $\lambda_4>0$, such that the FIMP is dominantly produced in the decay $H^\pm \rightarrow N_1\, \ell^\pm_{\alpha}$, where $H^\pm$ can be produced either directly in the partonic collision, or in the decay $H^0/A^0\rightarrow H^\pm W^\mp$. The dominant production channels for the charged scalar $H^\pm$ are shown in Fig.~\ref{Fig:SpecAHcProdProcesses}. We remark that in the following  we impose $\lambda_5 \ll 1$, therefore the decays $H^0/A^0\to A^0/H^0+Z^*$ are  suppressed, and we neglect them in our analysis.
The DY production cross-section of charged scalar pairs at the LHC, for a center-of-mass energy $\sqrt{s}=8$ TeV, is reported in Fig.~\ref{Fig:HcProdCMSlimit} for two benchmark scenarios, which are defined by the scalar mass splitting $\Delta m_{nc}\equiv m_{H^0/A^0}-m_{H^\pm}$, namely $\Delta m_{nc}=10$ GeV (solid black line) and $\Delta m_{nc}=70$ GeV (dashed black line).

 At the LHC the signals of the scotogenic FIMP scenario crucially depend on the masses of the $Z_2$-odd fermions relative to the $Z_2$-odd scalars. We consider here the following three representative scenarios:
\begin{itemize}
\item Scenario A: $M_1<m_{H^\pm, H^0, A^0}<M_{2,3}$  
\item Scenario B: $M_1<M_2<m_{H^\pm, H^0, A^0}<M_3$
\item Scenario C: $M_1<M_2<M_3<m_{H^\pm, H^0, A^0}$
\end{itemize}
with  the mass ordering in the scalar sector $m_{H^\pm}< m_{H^0}\simeq m_{A^0}$, which corresponds, as indicated earlier, to $\lambda_5 \ll 1$ and $\lambda_4>0$. Let us discuss each of them separately.

\mathversion{bold}
\subsection{Scenario A: $M_1<m_{H^\pm,H^0,\,A^0}<M_{2,3}$}
\mathversion{normal}\label{sub3.1}

In this scenario, the charged scalar $H^\pm$ can only decay into the FIMP and a charged lepton, with a rate given by \cite{Molinaro:2014lfa}
\begin{equation}
	\Gamma\left(H^{\pm}	\to N_{1} \,\ell^\pm_{\alpha}  \right)  =  \frac{m_{H^{\pm}}\,\left| Y^{\nu}_{\alpha 1} \right|^{2}}{16\,\pi}\left(1-\frac{M_{1}^{2}}{m_{H^{+}}^{2}}\right)^{2}\,\approx \frac{m_{H^{\pm}}\,\left| Y^{\nu}_{\alpha 1} \right|^{2}}{16\,\pi}\,.\label{eq:HtoNell}
\end{equation}
For FIMP dark matter, the proper decay-length of $H^\pm$, $c\tau (H^\pm)$, can be readily computed from Eqs.~(\ref{eq:HtoNell}) and (\ref{y1}). The result is
\begin{equation}
	c\tau(H^\pm)\;\approx\; 8.3\,\text{m}\,\left(\frac{M_1}{\text{10 keV}}\right)\left(\frac{\text{100 GeV}}{m_{H^\pm}}\right)^2\,\label{Lvac}
\end{equation}
 and  can clearly exceed the size of the ATLAS and CMS detectors.

The long decay-lengths expected for $H^\pm$ result in a charged track signal when this particle traverses the inner detector and the muon system. As the $H^\pm$ is very heavy compared to long-lived standard model particles, its momentum is  comparatively low and, as a consequence, the  amount of energy deposited in the detector is abnormally high. These features distinguish long-lived heavy charged particles from SM backgrounds, {\it i.e.} muons, thus allowing to define clean search regions  by vetoing velocities close to the speed of light and by requiring a high ionization. The CMS collaboration has conducted a search for heavy stable charged particles creating such tracks \cite{Chatrchyan:2013oca}, which was employed to derive constraints on the cross-section and the mass for selected benchmark scenarios. In Section \ref{sec:in-flight}, we recast this search taking into account that, for light dark matter, $H^\pm$ may not traverse the whole detector (as implicitly assumed by the CMS analysis), but may instead decay into a FIMP and a charged lepton. 

On the other hand, the charged scalars can be produced with such low momenta that they are stopped within the electromagnetic or hadronic calorimeters. Once trapped, they can decay at random times relatively to the trigger rate. When decaying during a time interval in which there is no bunch crossing, the signature can be a jet-like energy deposition that is largely free of backgrounds. The ATLAS collaboration conducted a search for stopped R-hadrons decaying out-of-time producing such signals \cite{Aad:2013gva}. Our study in Section \ref{StoppedHpm} uses the provided efficiency for a decay event to pass the out-of-time trigger of ATLAS, in order to assess the sensitivity to signals caused by stopped $H^\pm$.

\subsubsection{Charged-tracks analysis: in-flight decays of the $Z_2$-odd charged scalar}
\label{sec:in-flight}

\begin{figure}[t]
\begin{center}
\includegraphics[width=0.49\textwidth]{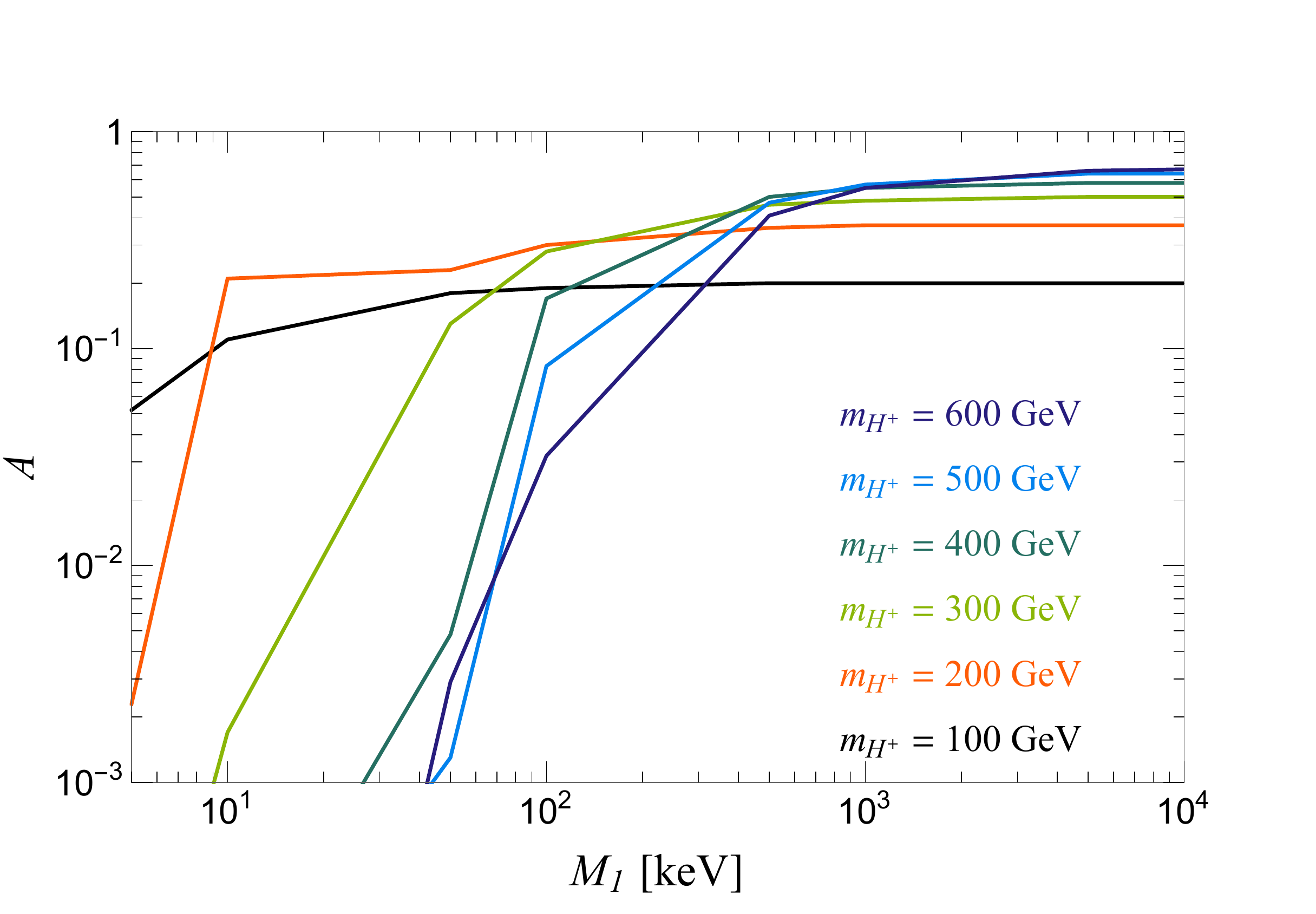} 
\includegraphics[width=0.49\textwidth]{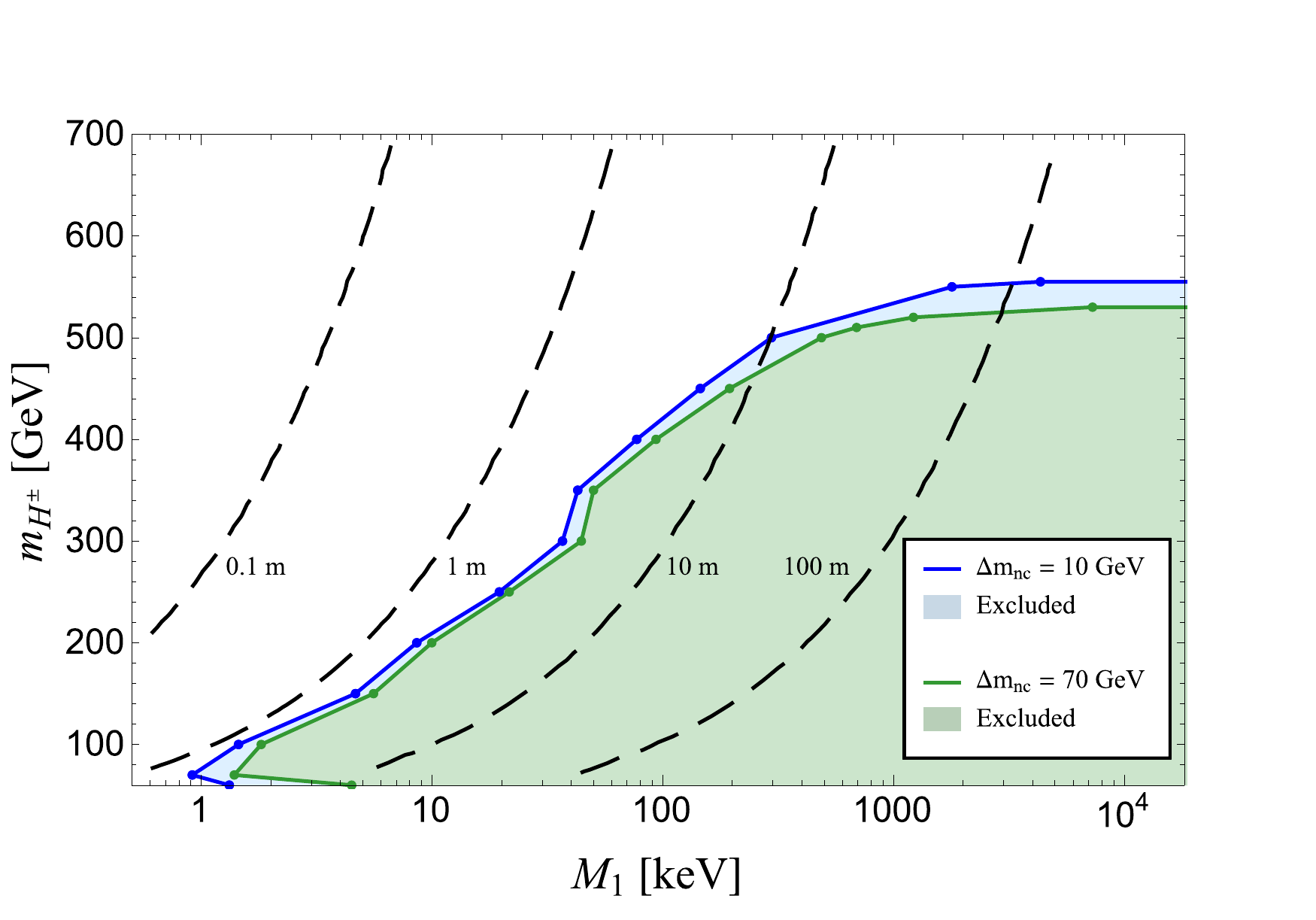}
\end{center}
\caption{\small{{\it Left panel}: Signal acceptance for the charged track analysis in Scenario A (see text for details), as a function of the FIMP mass for different values of the $Z_2$-odd charged scalar mass, $H^\pm$. {\it Right panel}: Region of the parameter space excluded at 95\% CL  in Scenario A for $\dmnc=10~(70)$ GeV;  the long-dashed contour lines correspond to different values of the  proper decay-length of $H^\pm$.}}
\label{Fig:Scenario-A}
\end{figure}

To derive constraints on the scotogenic FIMP model, we employ the null results from the search for metastable singly-charged particles leaving the detector conducted by the CMS collaboration  in \cite{Chatrchyan:2013oca}, based on $L=18.8$ fb$^{-1}$ of data collected at  $\sqrt{s}=8$ TeV.  The number of expected events reads
\begin{equation} \label{eq:exp-events-charged-track}
N_{\rm exp} =  \sigma \, L\,  \Acc\;,
\end{equation}
where  $\sigma $ is the production cross-section and $\Acc$ the signal acceptance.  We determine  $\Acc$  for a given $m_{H^\pm}$ and $M_1$ as  described in \cite{Khachatryan:2015lla} while taking  the finite particle decay-length fully into account. The acceptances are computed with a detector simulation of Monte Carlo events imposing the same kinematical cuts ($|\eta|<2.1$, $p_T>45$ GeV, $\beta<0.95$)  and isolation criteria as in \cite{Chatrchyan:2013oca}. In our treatment of the in-flight decays we approximate the geometry of the CMS detector by a barrel which covers the pseudorapidity range $0 \leq |\eta| < 1.2$  and has a radius $r=738$ cm, and endcaps covering $1.2 \leq \eta \leq 2.4$ which extend to $z=975$ cm from the production point. The acceptance we obtain for different values of $m_\Hpm$ is shown in Fig.~\ref{Fig:Scenario-A}, left panel, as a function of the FIMP mass for different values of $m_{H^\pm}$. As seen in the figure, the acceptance saturates for large $M_1$, as practically all produced charged $Z_2$-odd scalars leave the detector before decaying. 

We consider two benchmark scenarios for the mass spectrum in the $Z_2$-odd scalar sector, $\dmnc=10~(70)$ GeV, with Drell-Yan production cross-sections $\sigma$ given in Fig.~\ref{Fig:HcProdCMSlimit} (gluon-gluon fusion contributes negligibly to this signature). Finally, and following the procedure described in \cite{Khachatryan:2015lla}, we confront the expected number of events with the observed number of events in the search regions $m_{H^\pm}< 166\,{\rm GeV}$, $166 \,{\rm GeV}<m_{H^\pm}< 330\,{\rm GeV}$, $330 \,{\rm GeV}<m_{H^\pm}< 500\,{\rm GeV}$ and  $m_{H^\pm}> 500\,{\rm GeV}$. The regions of the parameter space spanned by $m_\Hpm$ and $M_1$ excluded by the CMS search are shown  in Fig.~\ref{Fig:Scenario-A}, right panel. As apparent from the figure, for low $M_1$ this search is practically insensitive to the scotogenic FIMP model, since a significant fraction of the charged $Z_2$-odd scalars decays inside the detector (contours of constant $c\tau$ are also shown in the plot, as long-dashed lines, for comparison). As $M_1$ increases, a larger and larger fraction of charged scalars leave the detector, thus strengthening the lower limit of $m_\Hpm$. The limit eventually saturates for $M_1\gtrsim 10$ MeV, when practically all the charged scalars are stable within the detector. In this regime, one finds the lower bound $m_{H^\pm} \; \gtrsim \; 560~(530)~\text{GeV}$ for $\Delta m_{\rm nc}=10~(70)~\text{GeV}$, regardless of the value of $M_1$.

\subsubsection{Decays of stopped long-lived $Z_2$-odd charged scalars}\label{StoppedHpm}

\begin{figure}[t!]
\begin{center}
\begin{tabular}{cc}
\includegraphics[width=0.5\textwidth]{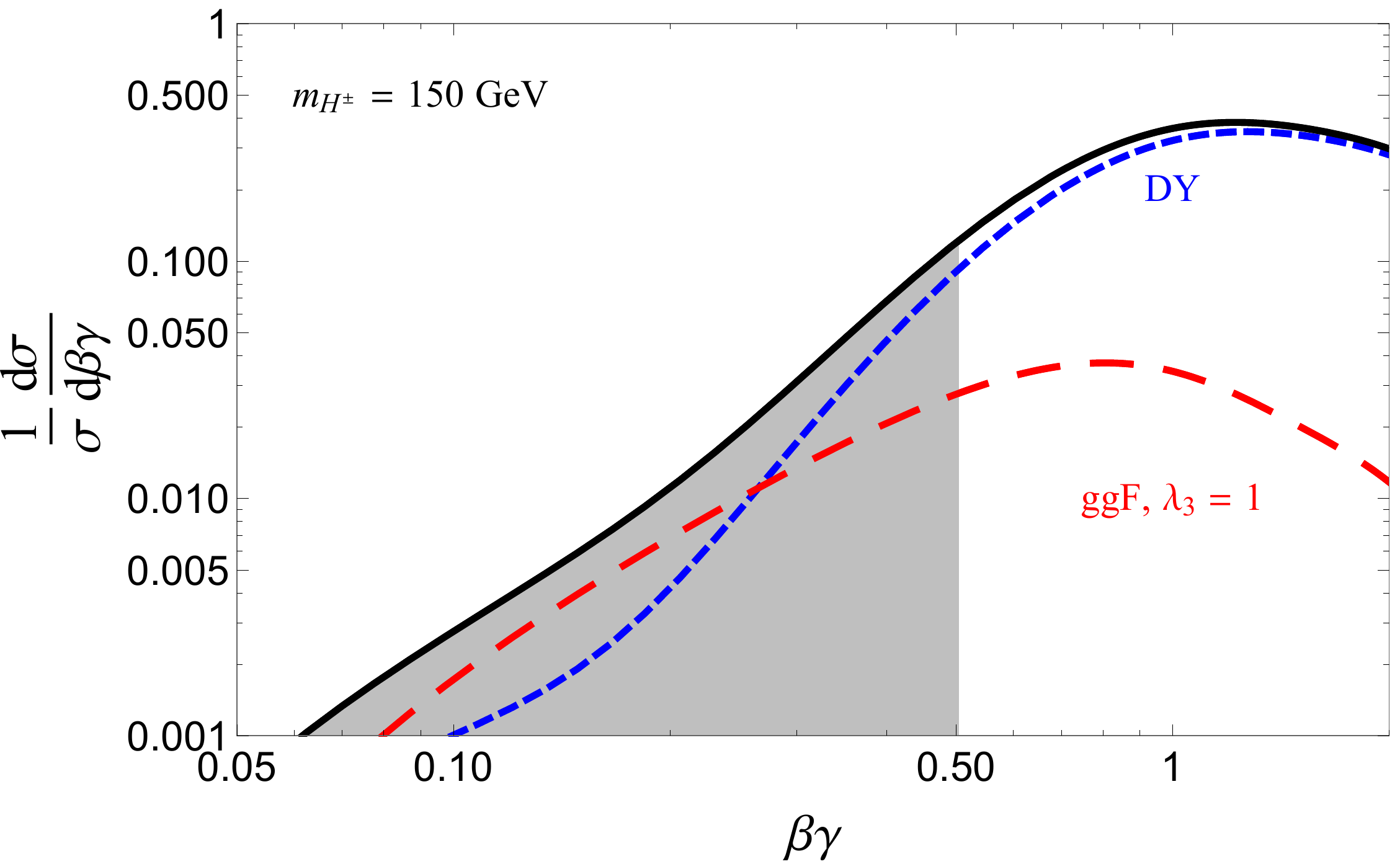} &
\includegraphics[width=0.5\textwidth]{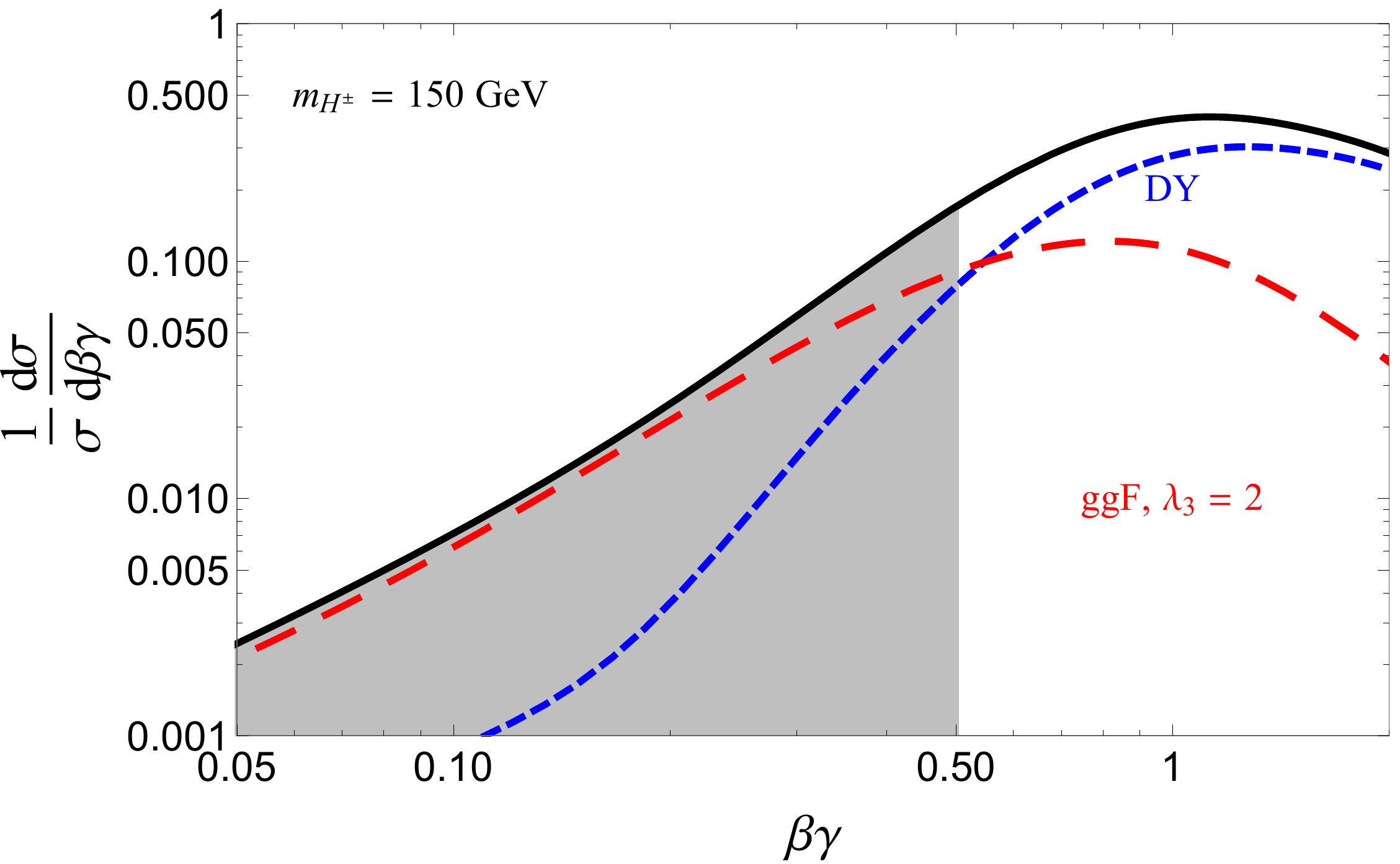} 
\end{tabular}
\caption{{\small Contribution to the $H^\pm$ distribution in $\beta\gamma$ from  Drell-Yan and gluon-gluon fusion at $\sqrt{s}=8$ TeV. 
We fix $m_{H^\pm}=150$ GeV and  $m_{H^0}=m_{A^0}=m_{H^\pm}+\dmnc$, with $\dmnc=10$ GeV.
The scalar quartic couplings in the left (right) panels are $\lambda_5=0$ and $\lambda_3=+1~(+2)$}.}
\label{Fig:distr}
\end{center}
\end{figure}

If the charged scalar is produced with sufficiently low momentum, it may be stopped in the detector before decaying, producing a characteristic signal. We report in Fig.~\ref{Fig:distr} the normalized differential cross section $\frac{1}{\sigma} \frac{d\sigma}{ d \beta \gamma}$  for  $H^\pm$  produced in proton-proton collisions at $\sqrt{s}=8$ TeV  for the benchmark mass  $m_{H^\pm}=150$ GeV,  assuming  $\dmnc=10$ GeV. We also show the fraction of particles which are produced from  DY and from ggF via an off-shell Higgs, through the Lagrangian term ${\cal L}\supset -\lambda_3\, v\, h \,H^\pm\, H^\mp$; the left panels show the results for $\lambda_3=1$ and the right panels for  $\lambda_3=2$.\footnote{We have verified that this set of parameters is in agreement with constraints from $h\to\gamma\gamma$ (to which $H^\pm$ contributes via a triangle loop), potential boundedness and unitarity of scalar scattering amplitudes.} The  distributions were obtained with CalcHEP \cite{Belyaev:2012qa}. 
The grey shaded region in each plot shows the $\beta\gamma$ values for which the charged scalar particle will stop within the barrel region ($|\eta|<1.2$) of the simplified ATLAS model employed in our analysis. The specifications of the detector are summarized in Tab.~\ref{Table:ToyDetATLAS} in Appendix~\ref{app:stopped}. As can be seen,  the ggF mechanism enhances the number of particles stopped within the detector  for some choices of the parameters of the model even though the ggF contribution to the total cross-section remains subleading.

Among all the charged scalars produced, only a fraction will eventually be stopped in the detector. We determine the stopping efficiency, $\epsilon_{\text{stop}}$, for different masses of $H^\pm$  
following the procedure detailed in Appendix~\ref{app:stopped}. For $\lambda_3=0$,  we find $\epsilon_{\text{stop}}=0.0076\;(0.0065)$ for $m_\Hpm =150$ GeV and  $\sqrt{s}=7$ TeV (8 TeV). For non-vanishing $\lambda_3$, the ggF channel increases the efficiency to $\epsilon_\mathrm{stop}=0.0095\;(0.0085)$ for a quartic coupling $\lambda_3=1$ and to $\epsilon_\mathrm{stop}=0.0129\;(0.0133)$ for $\lambda_3=2$.
 Once trapped, the charged scalars can decay at random times relative to the trigger rate. Searches for out-of-time decays are conducted in the time interval in which there is no bunch crossing. The timing acceptance  $\epsilon_T(\tau)$ depends on the operation mode of the LHC, in particular the bunch structure, and was presented in  Fig.~7 of \cite{Aad:2013gva}: $\epsilon_T(\tau)$ is identically zero for lifetimes $\tau\lesssim10^{-7}\,{\rm s}$ , then rises to a constant plateau of $\approx0.08$ starting at $\tau\approx 10^{-5}\,{\rm s}$ and starts to decrease at $\tau\approx 10^3\,{\rm s}$. This implies, in particular, that this search is only sensitive for FIMP masses $M_1\gtrsim1$ MeV when $m_\Hpm\sim 100$ GeV, since for smaller masses the charged scalar lifetime is smaller than $10^{-7}\,{\rm s}$, for which the timing efficiency is zero.

The decay products can then be detected. We use  the results of the search  for stopped R-hadrons decaying out-of-time conducted  by the ATLAS collaboration~\cite{Aad:2013gva}. This search relies on the observation of jets. Consequently, it is only sensitive to events of the type $H^\pm\to N_1\, \tau^\pm$ followed by $\tau^\pm\to$\;hadrons  and not to electrons or muons. Furthermore, the search is only sensitive to events with a leading jet energy $E > 50$ GeV. In order to estimate the effect of this cut, we simulate $\tau$ leptons from $H^\pm$ decays at rest with Pythia8~\cite{Sjostrand:2014zea}  and use Delphes~\cite{deFavereau:2013fsa} to include detector effects. For $m_\Hpm > 300$ GeV, the reconstruction efficiency approaches the limit imposed by the leptonic branching fraction of the $\tau$,  and reads  $\epsilon_\mathrm{rec}=0.65$. For lower masses, the efficiency substantially degrades, being $\epsilon_\mathrm{rec}=0.39$ (0.23) for $m_{H^\pm} =150$ (120) GeV.
The number of expected events then reads:
\begin{equation}
	N_{\rm exp} \; = \; \sigma\,L  \,\Estop\,\Erec\, \Etiming(\tau)\,.\label{Nobs2}
\end{equation}
Assuming ${\rm BR}(H^\pm\to\tau^\pm N_1)\approx 1$ and $\lambda_3=0$ we obtain $N_{\rm exp}=1.0$ ($0.60$) expected signal events for $m_{H^{\pm}} = 120$ GeV (150 GeV). For non-vanishing $\lambda_3$, the total production cross-section is enhanced by the ggF mechanism, and the number of expected signal events increases to $N_{\rm exp}=1.3$ ($0.78$) for $\lambda_3=1$ and to $N_{\rm exp}=2.0$ ($1.3$) for $\lambda_3=2$.

The ATLAS search \cite{Aad:2013gva} found an event rate in agreement with the background expectation.
Using the observed events and the background expectation in the search region with a leading jet energy larger than $50 $ GeV, one obtains, using the Feldman-Cousins procedure~\cite{Feldman:1997qc}, an  upper limit on the number of signal events of $N_{\rm sig} \lesssim 4.3$.  Due to the small expected signal, this search is therefore not sensitive enough to probe our scenario.

\mathversion{bold}
\subsection{Scenario B: $M_1<M_2<m_{H^0,\,A^0,\,H^\pm}<M_3$}
\mathversion{normal}

When $N_2$ is lighter than the $Z_2$-odd scalars, the decay modes $H^{\pm} \to N_2\, \ell^\pm_{\alpha}$ become kinematically allowed, with rate \cite{Molinaro:2014lfa}:
\begin{align}
	\Gamma\left(H^{\pm}\to N_2\, \ell^\pm_{\alpha} \right) = \frac{m_{H^{\pm}}\,\left| Y^{\nu}_{\alpha 2} \right|^{2}}{16\,\pi}\left(1-\frac{M_{2}^{2}}{m_{H^{\pm}}^{2}}\right)^{2}\,\approx \frac{m_{H^{\pm}}\,\left| Y^{\nu}_{\alpha 2} \right|^{2}}{16\,\pi} \;. \label{Hpmdecay}
\end{align}
Neutrino oscillation data and the upper limit on ${\rm BR}(\mu\rightarrow e\gamma)$ favour $y_2\equiv (\sum_\alpha |Y^{\nu}_{\alpha 2}|^2)^{1/2}$ in the range  $\sim 10^{-5}\div10^{-2}$, {\it cf.} Eq.~(\ref{boundy23}). Therefore $H^\pm$ will dominantly decay into the next-to-lightest $Z_2$-odd fermion, with a decay-length which is typically below 1~mm, while decays into $N_1$ are negligibly rare. 

In this scenario, instead, FIMPs are dominantly produced in the decay of the next-to-lightest $Z_2$-odd fermion, through  $N_2 \rightarrow \ell^-_{\alpha} \ell^+_{\beta} N_1$ and  $N_2 \rightarrow \nu_{\alpha} \bar{\nu}_{\beta} N_1$. The rates for these processes read:
\begin{align}
	\Gamma(N_{2} \to \ell^-_{\alpha} \ell^+_{\beta} N_1)\simeq 
	\Gamma(N_{2} \to  \nu_{\alpha} \bar{\nu}_{\beta} N_1)
	 \simeq  \frac{M_{2}^{5}}{6144\,\pi^{3}\,m_{H^{\pm}}^{4}}\left(\left|Y^{\nu}_{\beta 1} \right|^{2} \left|Y^{\nu}_{\alpha 2} \right|^{2}+\left|Y^{\nu}_{\alpha 1} \right|^{2} \left|Y^{\nu}_{\beta 2} \right|^{2}\right)\,,\label{N2decay}
\end{align}
For the values of the FIMP coupling to the leptons required to correctly reproduce the observed dark matter abundance via freeze-in, Eq.~(\ref{y1}), the decay-length in vacuum of $N_2$ is:
\begin{align}
  c\tau(N_2)\;\approx\; 2\times 10^{13}\,\text{m}\,\left(\frac{M_1}{\text{10 keV}}\right)\left(\frac{m_{H}}{\text{500 GeV}}\right)^3\left(\frac{\text{100 GeV}}{M_2}\right)^5\left(\frac{10^{-3}}{y_2}\right)^2
  \label{eq:N2toN1}
\end{align}
where we take $m_H=m_{H^\pm}\approx m_{H^0}$. This decay-length is orders of magnitude larger than the size of the detector and can not be probed at the moment.\footnote{This could change in the future if the recently proposed MATHUSLA surface detector~\cite{Chou:2016lxi}, which has the potential to add  sensitivity to particles with  $c\tau \gtrsim 10$ m to the existing LHC program, is built.}
  Therefore, the experimental signal of Scenario B consists in the observation of two prompt charged leptons and missing energy, from the production of a $H^+ H^-$ pair, followed by the decay $H^\pm\to N_2\, \ell_\alpha^\pm$.  A similar signature arises in simplified models of Supersymmetry with light sleptons and weakly-decaying charginos, which has been searched for by the ATLAS collaboration in  \cite{Aad:2014vma}. We simulate the production and decay of pairs of the  scalars $H^{\pm}/A^0/H^0$ with  CalcHEP~\cite{Belyaev:2012qa}  and pass the result to Pythia~\cite{Sjostrand:2014zea} for showering and hadronization. Finally, we use the recast of the experimental analysis implemented in Checkmate~\cite{Drees:2013wra} which uses Delphes~\cite{deFavereau:2013fsa} to simulate detector effects.~\footnote{In Checkmate the ATLAS search \cite{Aad:2014vma} is available as an unvalidated analysis.  To check the implementation we re-derived the limits on slepton production and find good agreement between the Checkmate result and the experimental analysis.}

\begin{figure}[t]
\begin{center}
\includegraphics[width=0.5\textwidth]{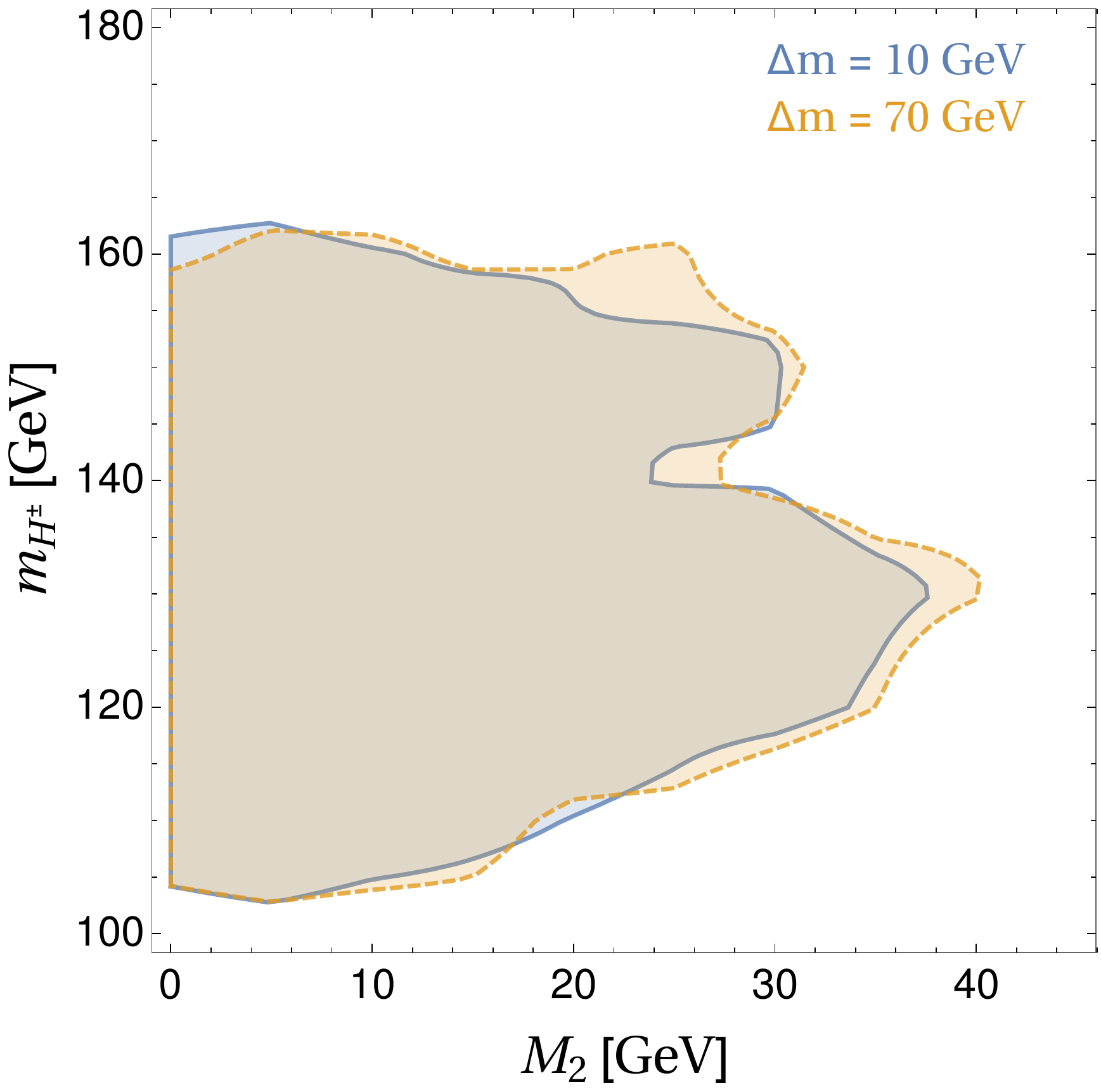} 
\caption{{\small 
Parameter space of the Scenario B excluded by the ATLAS search for dileptons and missing energy \cite{Aad:2014vma} for $\Delta m_{\rm nc} = 10 \;  \mbox{GeV}$ (blue) and $\Delta m_{\rm nc} =  70 \; \mbox{GeV}$ (yellow), assuming   $ {\rm BR}(H^\pm  \rightarrow N_2\,\mu^\pm)={\rm BR}(H^\pm  \rightarrow N_2\,e^\pm) =1/2$ and $ {\rm BR}(H^\pm  \rightarrow N_2\,\tau^\pm) =0$. }}
\label{Fig:SpecB}
\end{center}
\end{figure}

The analysis presented in \cite{Aad:2014vma} considers only final states with muons or electrons. As a result, the exclusion limits will show a strong dependence on the branching ratio into taus. We then consider two cases: $i)$ an optimistic scenario in which the decays into taus are negligible and $ii)$ a more conservative benchmark with $\mbox{BR}(H^{\pm} \rightarrow N_2\, e^\pm)=\mbox{BR}(H^{\pm} \rightarrow  N_2\,\mu^\pm)=\mbox{BR}(H^{\pm} \rightarrow  N_2\,\tau^\pm)=1/3$. As can be seen in Fig.~\ref{Fig:SpecB}, for the optimistic case the LHC  probes a significant region of the parameter space and is able to exclude $m_{H^\pm} \lesssim 160 $ GeV for light  $N_2$. In contrast, we find  that once $\mbox{BR}(H^{\pm} \rightarrow N_2\,  \tau^\pm)$ is equal or larger than the branching ratio into electrons and muons, the whole parameter space becomes allowed.

\mathversion{bold}
\subsection{Scenario C: $M_1<M_2<M_3<m_{H^0,\,A^0,\,H^\pm}$}
\mathversion{normal}

The scenario where all the $Z_2$-odd fermions are lighter than the $Z_2$-odd scalars produces two different signatures. The first signature arises from the decay $H^\pm \rightarrow N_{2} \,\ell_\alpha^\pm$, which produces two charged leptons plus missing energy and, as $N_2$ has a long decay-length, this signal is identical to the one already discussed in Scenario B. The second signature arises from the decay $H^\pm \rightarrow N_{3}\, \ell_\alpha^\pm$. In contrast to $N_2$ which is stable within the detector, $N_3$ can decay fast enough into $\ell_{\alpha} \,\bar{\ell}_{\beta}\, N_2$ to produce observable signals. The decay rate reads, assuming $M_2\ll M_3$,
\begin{align}
	\Gamma(N_{3} \to \ell^-_{\alpha} \ell^+_{\beta} N_2)
	 \simeq  \frac{M_{3}^{5}}{6144\,\pi^{3}\,m_{H^\pm}^{4}}\left(\left|Y^{\nu}_{\beta 2} \right|^{2} \left|Y^{\nu}_{\alpha 3} \right|^{2}+\left|Y^{\nu}_{\alpha 2} \right|^{2} \left|Y^{\nu}_{\beta 3} \right|^{2}\right)\,,\label{N2decay}
\end{align}
Taking $m_H=m_{H^\pm}\approx m_{H^0}$ again,  the proper decay-length of $N_3$ is
\begin{align}
  c\tau(N_3)\;\approx\; 0.4\,\text{m}\left(\frac{\text{100 GeV}}{M_3}\right) \left(\frac{m_{H}}{M_3}\right)^4\left(\frac{10^{-3}}{y_2}\right)^2\left(\frac{10^{-3}}{y_3}\right)^2
  \label{eq:N3toN2}
\end{align}
and can be macroscopic for some choices of the parameters of the model. Such displaced dilepton pairs are a very clean observable and can be searched for very efficiently at the LHC, see e.g.~\cite{Khachatryan:2014mea,CMS:2014hka}. Since limits on electron-positron pairs are typically a factor of a few weaker than those related to muons,  we focus on the search for displaced dimuons  presented in  \cite{CMS:2014hka} and recast the limits reported by CMS to  our model. For a description of the dilepton search and our treatment of the detector performance, we refer the reader to Appendix~\ref{app:dileptons}.

\begin{figure}[t]
\begin{center}
\includegraphics[width=0.6\textwidth]{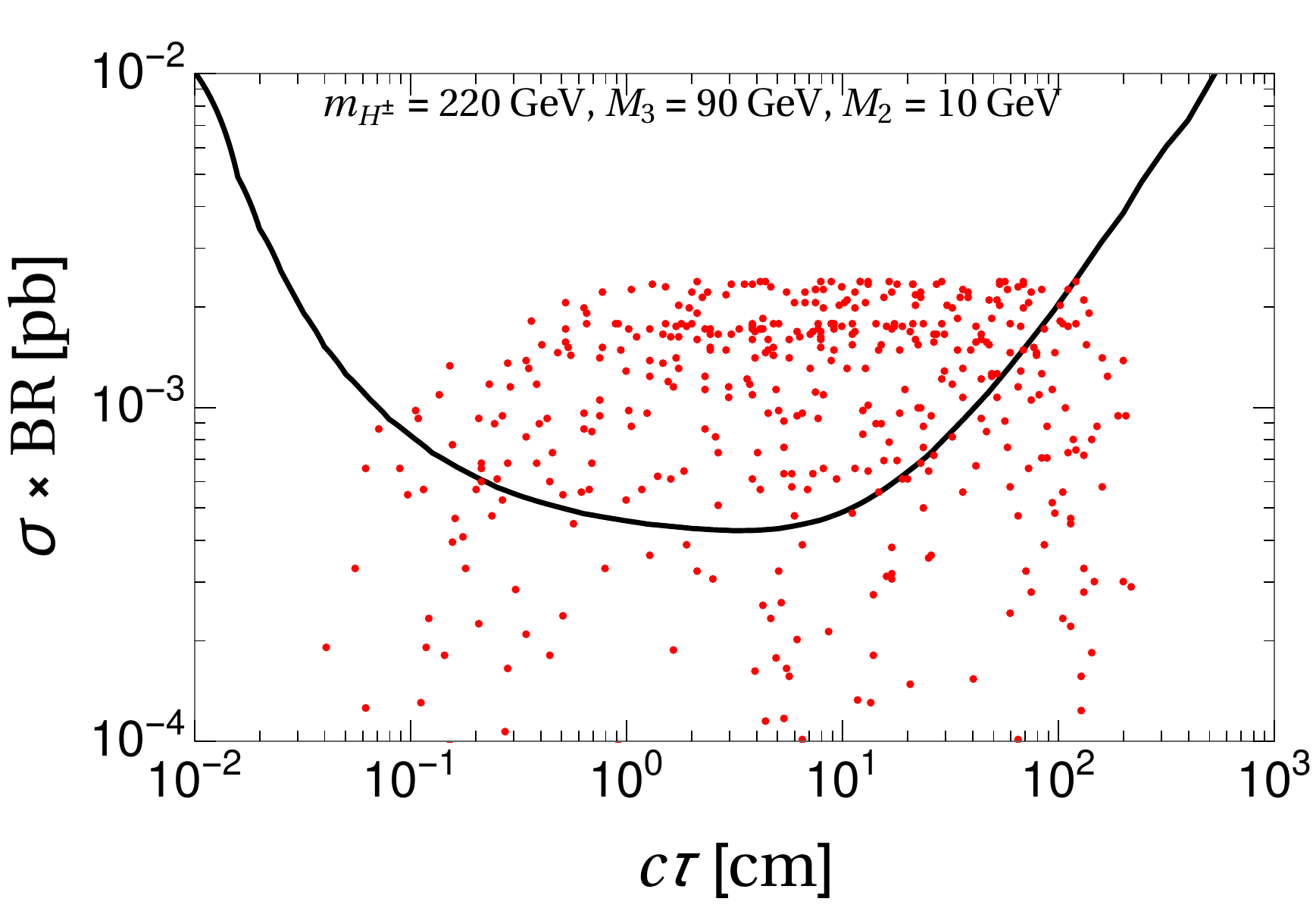} 
\caption{\small Cross-section times branching ratio into the $\mu^+ \mu^-$ channel as a   function of the proper decay-length $c\tau (N_3)$ for one exemplary choices of the scalar and right-handed neutrino masses in Scenario C. The solid black line corresponds to the $95\%$ CL exclusion derived from the CMS limits on displaced dileptons. }
\label{fig:ctaudimuon}
\end{center}
\end{figure}

The sensitivity of the experimental search is dictated by the dimuon production rate, which in turn depends on the $H^\pm$ production cross-section times the branching ratio ${\rm BR}(N_3\rightarrow \mu^+\mu^- N_2)$, and by the displacement of the dimuon vertex from the collision point, which depends on the $N_3$ decay-length. We show  in Fig.~\ref{fig:ctaudimuon} the limits on the cross-section times branching ratio as a function of the proper decay-length $c \tau$ that we obtain from recasting the CMS results to this scenario. The excluded cross-section is fairly insensitive to the precise value of the proper decay-length in the  range $0.1 \-- 10$ cm, while for larger and smaller values there is a rapid decrease of the sensitivity, since only a small fraction of the particles will decay in the tracker.
To assess the impact of this search on the FIMP scotogenic model, we also show in the plot the predicted values of $\sigma\times {\rm BR}$ and $c\tau$ obtained from a random scan of the Yukawa couplings of the model, and fixing for illustration $m_{H^\pm}=220\,{\rm GeV}$, $M_3=90\,{\rm GeV}$ and $M_2=10\,{\rm GeV}$. The Yukawa couplings were constructed, following the approach introduced in \cite{Casas:2001sr}, such that they lead to the observed neutrino mass splittings and mixing angles. Furthermore, it was checked that all the points in the scan are in agreement with the upper limit on ${\rm BR}(\mu\rightarrow e\gamma)$. As seen from the plot, a significant fraction of the parameter space of the model can be probed in searches for displaced muon pairs.

\begin{figure}[t]
\begin{center}
\includegraphics[width=0.49\textwidth]{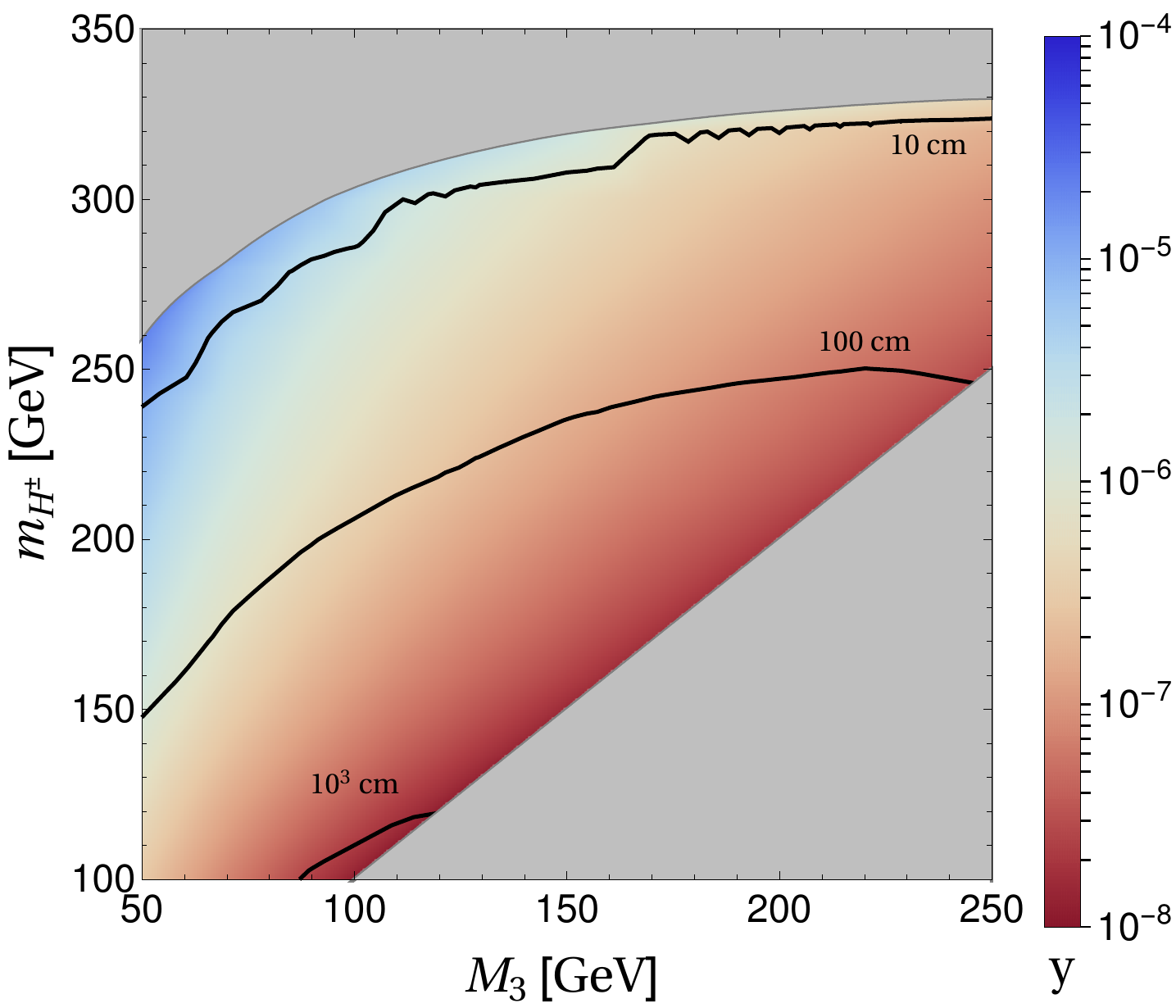} 
\includegraphics[width=0.49\textwidth]{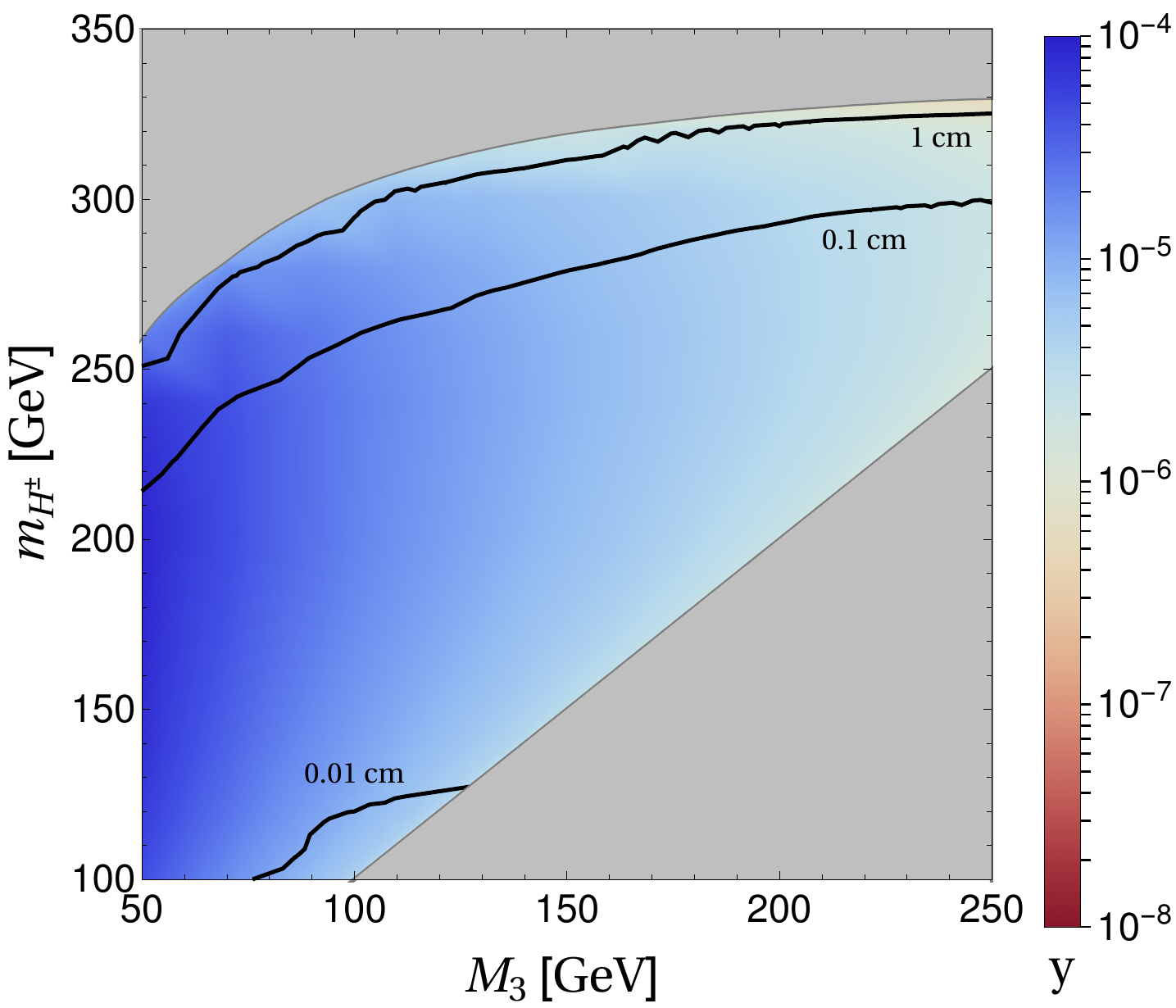} 
\caption{\small Smallest (left panel) and largest (right panel) combination of Yukawa couplings $y$ excluded by current displaced dimuon searches at CMS in Scenario C, assuming ${\rm BR}(N_3\rightarrow \mu^+\mu^- N_2)=1/18$ and $M_2=10 \;{\rm GeV}$. The black contour lines indicate the excluded proper decay-length $c\tau$. In the upper grey region the production rate is too low to allow an exclusion for any decay-length, whereas the lower grey region is kinematically inaccessible.}
\label{fig:limits-C}
\end{center}
\end{figure}

Conversely, one can use the CMS null results to constrain the fundamental parameters of the model relevant for this search, which are $m_{H^\pm}$, $M_3$, $M_2$ and $|Y^\nu_{\mu 2}Y^\nu_{\mu 3}|$. Under the simplifying assumption $M_2\ll M_3$ and fixing ${\rm BR}(N_3\rightarrow \mu^+\mu^- N_2)=1/18$, which  is a reasonable expectation for an anarchic Yukawa matrix with similar couplings to all flavours, the free parameters of the model reduce to $m_{H^\pm}$, $M_3$ and $y\equiv \sqrt{|Y^\nu_{\mu 2}Y^\nu_{\mu 3}|}$. It is easy to see that, for fixed values of $m_{H^\pm}$ and $M_3$, the CMS search can set a lower and an upper limit on $y$: for smaller values of $y$, $N_3$ becomes long-lived compared to the detector scales, while for larger values, $N_3$ decays too fast and does not leave a displaced dimuon signal. The lower and upper limits of $y$ in the $M_3$--$m_{H^\pm}$ plane are shown in Fig.~\ref{fig:limits-C}. The black lines in the plots represent the limit on the proper decay-length of $N_3$.

\section{Conclusions}
\label{sec:Conclusions}

We analysed  a scenario where the dark matter candidate is the lightest fermion singlet of the scotogenic neutrino mass model, under the assumption that the present dark matter abundance is generated via the freeze-in mechanism. The small interaction strength of $N_1$ required to correctly reproduce the observed dark matter abundance implies very suppressed decay rates of the next-to-lightest $Z_2$-odd particle, which translate into macroscopic decay-lengths.

We have investigated possible signals of this model in three representative scenarios, characterized by the spectra of the $Z_2$-odd particles. Each spectrum leads to very characteristic signatures at the Large Hadron Collider, which lead to different constraints on the model parameters:
\begin{itemize}
\item Scenario A: $M_1<m_{H^\pm, H^0, A^0}<M_{2,3}$. The  charged component of the $Z_2$-odd scalar doublet, $H^\pm$, is long-lived and leaves a highly ionizing charged track  in the detector. For $M_1\gtrsim 1$ MeV, most of the produced charged scalars leave the detector, producing a long track; the non-observation of this signal at the LHC excludes  $m_{H^\pm} \lesssim 500 $ GeV. On the other hand, for lighter dark matter, the proper decay-length becomes comparable or smaller than the size of the detector. In this case, the sensitivity of the search decreases. Nevertheless, we still find significant constraints on the parameter space, namely we find $m_{H^\pm} \gtrsim 400 $ GeV (200 GeV) for $M_1=100$ keV (10 keV). For $M_1\lesssim 1$ keV, the charged scalar decays fast and does not leave any observable track. Furthermore, for $M_1\gtrsim 1$ MeV, a fraction of the produced charged scalars is stopped in the detector and may decay in the time interval in which there is no bunch crossing. We find that, unfortunately, current limits on this class of exotic events are not strong enough to constrain this model.
\item Scenario B: $M_1<M_2<m_{H^\pm, H^0, A^0}<M_3$. In this scenario, the charged scalar decays promptly to $N_2$ and a charged lepton, producing a signature consisting of hard leptons and missing energy. We find that, when $N_2$ couples mostly to the electron or the muon flavour, charged scalar masses up to  $\approx 160$~GeV can be excluded.
\item Scenario C: $M_1<M_2<M_3<m_{H^\pm, H^0, A^0}$. In this case, the non-observation of the process $\mu\rightarrow e\gamma$ translates into small Yukawa couplings of $N_3$ and $N_2$ to the charged leptons, which in turn translates into a rather small width for the decay $N_3\rightarrow \ell^-_\alpha \,\ell^+_\beta \,N_2$. As a result, this scenario can be probed by searching for displaced dileptons at the LHC. Notably, and in contrast to the other two scenarios, if this one is realized in Nature, it may be possible to directly probe the parameters responsible for neutrino masses. We find that current searches for displaced dimuons at CMS already probe regions of the parameter space allowed by other searches and set quite stringent constraints on the size of the Yukawa couplings.
\end{itemize}

The FIMP realization of the scotogenic model presents a rich phenomenology which can be probed using various searches for long-lived particles at the LHC. The model could then serve as a proxy to assess the impact of these searches for concrete Particle Physics models. Given the possible relevance of this searches for understanding the nature of the dark matter and the origin of neutrino masses, we would like to encourage the experimental collaborations to strengthen their efforts in conducting searches for long-lived particles, especially in association with energetic leptons in the final state. Finally, it should be noted that new detectors specifically designed to address the problems posed by long-lived particles, for example the proposed MATHUSLA experiment~\cite{Chou:2016lxi}, could extend the sensitivity well beyond the range currently probed at the LHC.

\section{Acknowledgements}

We would like to thank Moritz Platscher for helpful comments. This work was supported in part by the DFG cluster of excellence EXC 153 ``Origin and Structure of the Universe''. 
The CP3-Origins center is partially funded by the Danish National Research Foundation, grant number DNRF90. A. G. H. was supported by Funda\c c\~ao para a Ci\^encia e a Tecnologia (FCT) through the grant SFRH/BD/76052/2011, financed by the European Social Fund (ESF) through POPH under the QREN framework.

\appendix

\appendix

\section{Stopped particles}
\label{app:stopped}

The stopped particles analysis requires a realistic  description of the ATLAS detector and the energy losses therein, which depend crucially on the material properties. The mean energy loss of particles in a given material is described by the Bethe-Bloch equation \cite{Agashe:2014kda},
\begin{align}\label{Eq:BetheBloch[E,x,eta]}
\dedxmeanminus=&\,\kappa\ssp z^2\ssp\ZAmean\ssp\rho\ssp\frac{1}{\beta^2}\left[\ln\left(\frac{2\ssp \,m_e}{I}\frac{E^2-M^2}{M\sqrt{M^2+2\,m_e\,E+m_e^2}}\right)-\beta^2\right]\,,
\end{align}
where $E$ is the energy, $M$ is the mass, $z$ the electric charge and $\beta=v/c$, with $v$ being the velocity of the particle. The constant $\kappa$ has the value $0.307$ MeV\,g$^{-1}$\,cm$^2$. The quantities $\rho$, $I$, $Z$ and $A$,  describe the material and represent the density, mean excitation energy, atomic number and atomic weight of the medium, respectively. $\ZAmean$ is the mean weighted over isotopic abundances on Earth.  
These quantities are not constant throughout the detector but vary along the path of the particles. Given that the experimental search~\cite{Aad:2013gva} only considers events with $|\eta|< 1.2$ it is sufficient to consider the barrel.  In the following we briefly describe the simplified detector used in our analysis. A summary of the detector geometry and the material constants can be found  in Tab.~\ref{Table:ToyDetATLAS}~\footnote{The information on geometry and material properties was extracted from the official technical descriptions of the ATLAS experiment \cite{Aad:2008zzm} and from publicly available technical design reports (TDRs) \cite{ATLAS:1996ab,ATLAS:1996aa,ATLAS:1997ad}. Numerical values for the material constants are taken from the PDG \cite{Agashe:2014kda}, unless otherwise specified in the TDRs.}.

To incorporate the cylindrical geometry, we define the material constants in terms of box functions of the distance from the interaction point $x$ and of the pseudorapidity $\eta$.  The solenoid (Sol.), the muon system (MS) and the toroid (Tor.) of ATLAS have a low density and we neglect them in our treatment of the energy losses.  Significant energy losses occur in the electromagnetic calorimeter (ECAL) and the hadron calorimeter (HCAL) composed by the long (HLB) and extended barrel (HEB) modules in the region of interest. Particles which pass the calorimeters will not contribute to the stopped particle signal. The calorimeters are not homogeneous slabs of material and, therefore, the  material coefficients entering the Bethe-Bloch equation need to be averaged over the different components. For compound materials, we use the Bragg additivity rule \cite{Groom:2001kq},~i.e. we sum the different $\log(I)$ and $Z/A$ of the various materials weighted by their respective mass fraction.

The ECAL barrel, which is part of the liquid argon (LAr) calorimeter system, is subdivided into two regions with different LAr and lead (Pb) proportions; these pseudorapidity ranges are denoted by R1 and R2 in Tab.~\ref{Table:ToyDetATLAS}. The geometry and composition of the various ECAL layers is not simple enough to allow for an easy determination of an effective average density. ATLAS sources \cite{Aurousseau:2010oaa,Abdelalim:2008gsa} apply a method with an ``effective molecule" with numbers of representative atoms chosen to reproduce the overall proportions of lead, reinforcing steel and argon.  For the ECAL barrel in the region R1, this is stated to yield a density of 4.01 g/cm$^{3}$. We use the same molecule combined with Bragg additivity to compute the effective density for the other region, with atom numbers in the chemical formula adjusted by factors obtained from the different thicknesses of the absorber and radiator layers. Moreover, we also apply the Bragg rule to compute the means of $I$ and $Z/A$. In the HCAL, $82\%$ of the volume is made up by iron while the rest is taken by plastic scintillators. We once again use the  Bragg additivity rule to perform the material averages.

With these ingredients we can solve the Bethe-Bloch equation and, since energy losses are treated as continuous, we get, for each direction in the detector, a maximal value of $\beta$ for which the charged particles stop.
The stopping efficiency $\Estop$ is then given by
\begin{equation}
\Estop(\tau)=\int_{|\eta|<1.2}d\eta\int_1^{\beta\gamma_\mathrm{lim}(\eta)}d\beta\gamma\,\frac{1}{\sigma}\frac{d^2\sigma}{ d \beta \gamma\, d \eta}\,P_\mathrm{sur}(\beta\gamma,\eta)\;.
\end{equation}
We perform the integration over the pseudorapidity range considered in the experimental analysis and up to the maximum value $\beta\gamma_\mathrm{lim}(\eta)$ at the point of production for which the particle with pseudorapidity $\eta$ gets trapped in the ECAL or the HCAL. This value is given by the solution of the Bethe-Bloch equation. Lastly, $P_\mathrm{sur}$ accounts for the instability of the $\Hpm$ particles, namely it corresponds to the probability to survive until the particle is stopped. However, this factor is only relevant for very short lifetimes which can not be probed at the LHC due to the vanishing timing efficiency $\epsilon_T(\tau)$, hence we take $P_\mathrm{sur}=1$ in our analysis.

\begin{table}[t]
\begin{center}
\begin{small}
\begin{tabular}{l|ccccccc}
Layer            & $\Delta$ [cm] & $[|\eta|_\mathrm{min},\,|\eta|_\mathrm{max}]$ & Material & $\rhomean$ [g/cm$^3$] & $\Imean$ [eV] & $\ZAmean$\\
\Xhline{2.2\arrayrulewidth}
ID+Sol.          & 150.0          & $[0,\,1.4]$     & Vacuum    & -     & -     & -     \\
EMB R1           & 47.6           & $[0,\,0.8]$     & LAr+Pb    & 4.01  & 487   & 0.406 \\
EMB R2           & 47.6           & $[0.8,\,1.4]$   & LAr+Pb    & 3.67  & 447   & 0.408 \\
HLB+HEB          & 197.0          & $[0,\,1.4]$     & Fe+PS     & 6.40  & 286   & 0.466 \\
MS+Tor.          & $\infty$       & $[0,\,1]$       & Vacuum    & -     & -     & -     
\end{tabular}
\end{small}
\end{center}
\caption{\small{
Specs of our simplified ATLAS barrel. See the text for details.
}}
\label{Table:ToyDetATLAS}
\end{table}

\section{Displaced dileptons}\label{app:dileptons}

The CMS  displaced dilepton analysis \cite{CMS:2014hka} searches for pairs of electrons or muons originating from a secondary vertex with a substantial separation from the collision point.
As the limits on muon pairs are considerably stronger than the ones for electrons we will focus our attention on the muon channel. 
The  detector model used in our analysis follows roughly the phenomenological recast described in \cite{Liu:2015bma}.
 
The first ingredient for a reinterpretation of the CMS analysis is a description of the detector and the efficiencies with which tracks originating  from displaced vertices can be observed.
Unfortunately, the  performance of the CMS tracker is not publicly  known in detail and we have to work with some simplified approximations. In the following we assume that the 
tracker is close to perfect and that the tracking efficiency depends exclusively on the 
position of the displaced vertex. Motivated by Fig.~3 of $\cite{CMS:2014hka}$  we model the
dependence of the  tracker efficiency on the transverse displacement $|d_0|$ as 
a broken linear function. Our  efficiency starts at 1 for $d_0$=0,
 falls to 0.8 at $|d_0|=15$ cm 
 and drops to 0 for $|d_0| \geq 30$. We take a similar approach for  the longitudinal displacement $|z_0|$. Here the efficiency falls to 0.8 at $|z_0|= 30$ cm before dropping to 0 for $|z_0|\geq  55 $ cm. Following  \cite{Liu:2015bma} we require a hard cut on the radial displacement $r$  and set the tracking efficiency to zero for $r \geq 60$ cm.

Once the probability for the detection of a lepton pair has been determined, we have to make sure that the event passes the experimental cuts for  dilepton candidates.
Both muons must have a $p_T > 26$~GeV in order to be sufficiently above the trigger threshold of $23$~GeV. 
The pseudorapidity of $|\eta| < 2$ is required in order to ensure that the leptons are observed in the well-instrumented region of the tracker.  A cut on the  total invariant mass of the lepton pair $m_{\ell \ell} >15$ GeV suppresses a possible contamination from meson decays. In addition, a  minimal separation of the two muon tracks of $\Delta R > 0.2$ is required to ensure  a high dimuon trigger efficiency. 
The absolute difference in azimuthal angle $|\Delta \Phi|$ between the displaced vertex and the momentum of the dilepton has to be smaller than $\pi /2 $. Finally, the tracks have to exhibit a significant transverse displacement $|d_0|$ from the primary vertex of more than $12 \sigma$, where $\sigma$ is the uncertainty of $d_0$. 
Since this uncertainty is not accessible  without a full detector simulation we replace this condition with a minimal displacement $|d_0|> 250 \,\mu\mbox{m}$.
The expected number of background events in this search is zero and, since no events were observed,  $N=3$ is  excluded at $95\%$ CL. 

In order to check the validity of our analysis we have rederived the CMS limits for some of  the representative benchmark scenarios presented in   \cite{CMS:2014hka}; our results agree with those from CMS within $30\%$.


\begin{thebibliography}{99}


\bibitem{Bertone:2010zza}
  G.~Bertone {\it et al.},
  Cambridge, UK: Univ. Pr. (2010) 738 p

\bibitem{Bergstrom:2000pn}
  L.~Bergstr\"om,
  Rept.\ Prog.\ Phys.\  {\bf 63} (2000) 793
  doi:10.1088/0034-4885/63/5/2r3
  [hep-ph/0002126].



\bibitem{Bertone:2004pz}
  G.~Bertone, D.~Hooper and J.~Silk,
  Phys.\ Rept.\  {\bf 405} (2005) 279
  doi:10.1016/j.physrep.2004.08.031
  [hep-ph/0404175].

\bibitem{Baer:2014eja}
  H.~Baer, K.~Y.~Choi, J.~E.~Kim and L.~Roszkowski,
  Phys.\ Rept.\  {\bf 555} (2015) 1
  doi:10.1016/j.physrep.2014.10.002
  [arXiv:1407.0017 [hep-ph]].

\bibitem{Gershtein:1966gg}
  S.~S.~Gershtein and Y.~B.~Zeldovich,
  JETP Lett.\  {\bf 4} (1966) 120
   [Pisma Zh.\ Eksp.\ Teor.\ Fiz.\  {\bf 4} (1966) 174].

\bibitem{Ade:2013zuv}
  P.~A.~R.~Ade {\it et al.} [Planck Collaboration],
  Astron.\ Astrophys.\  {\bf 571} (2014) A16
  doi:10.1051/0004-6361/201321591
  [arXiv:1303.5076 [astro-ph.CO]].


\bibitem{Hall:2009bx}
  L.~J.~Hall, K.~Jedamzik, J.~March-Russell and S.~M.~West,
  JHEP {\bf 1003} (2010) 080
  doi:10.1007/JHEP03(2010)080
  [arXiv:0911.1120 [hep-ph]].


\bibitem{Ma:2006km}
  E.~Ma,
  Phys.\ Rev.\ D {\bf 73} (2006) 077301
  doi:10.1103/PhysRevD.73.077301
  [hep-ph/0601225].
  
\bibitem{Ma:2006fn}
  E.~Ma,
  Mod.\ Phys.\ Lett.\ A {\bf 21} (2006) 1777
  doi:10.1142/S0217732306021141
  [hep-ph/0605180].
  
\bibitem{Kubo:2006yx}
  J.~Kubo, E.~Ma and D.~Suematsu,
  Phys.\ Lett.\ B {\bf 642} (2006) 18
  doi:10.1016/j.physletb.2006.08.085
  [hep-ph/0604114].
  
  

\bibitem{Molinaro:2014lfa}
  E.~Molinaro, C.~E.~Yaguna and O.~Zapata,
  JCAP {\bf 1407} (2014) 015
  doi:10.1088/1475-7516/2014/07/015
  [arXiv:1405.1259 [hep-ph]].


\bibitem{Aad:2012tfa}
  G.~Aad {\it et al.} [ATLAS Collaboration],
  Phys.\ Lett.\ B {\bf 716} (2012) 1
  doi:10.1016/j.physletb.2012.08.020
  [arXiv:1207.7214 [hep-ex]].


\bibitem{Chatrchyan:2012ufa}
  S.~Chatrchyan {\it et al.} [CMS Collaboration],
  Phys.\ Lett.\ B {\bf 716} (2012) 30
  doi:10.1016/j.physletb.2012.08.021
  [arXiv:1207.7235 [hep-ex]].


\bibitem{Hambye:2009pw}
  T.~Hambye, F.-S.~Ling, L.~Lopez Honorez and J.~Rocher,
  JHEP {\bf 0907} (2009) 090
   Erratum: [JHEP {\bf 1005} (2010) 066]
  doi:10.1007/JHEP05(2010)066, 10.1088/1126-6708/2009/07/090
  [arXiv:0903.4010 [hep-ph]].


\bibitem{Ginzburg:2003fe}
  I.~F.~Ginzburg and I.~P.~Ivanov,
  hep-ph/0312374.

\bibitem{Merle:2015ica}
  A.~Merle and M.~Platscher,
  JHEP {\bf 1511} (2015) 148
  doi:10.1007/JHEP11(2015)148
  [arXiv:1507.06314 [hep-ph]].

\bibitem{Toma:2013zsa}
  T.~Toma and A.~Vicente,
  JHEP {\bf 1401} (2014) 160
  doi:10.1007/JHEP01(2014)160
  [arXiv:1312.2840, arXiv:1312.2840 [hep-ph]].


  \bibitem{TheMEG:2016wtm}
  A.~M.~Baldini {\it et al.} [MEG Collaboration],
  Eur.\ Phys.\ J.\ C {\bf 76} (2016) no.8,  434
  doi:10.1140/epjc/s10052-016-4271-x
  [arXiv:1605.05081 [hep-ex]].

\bibitem{Ibarra:2016dlb}
  A.~Ibarra, C.~E.~Yaguna and O.~Zapata,
  Phys.\ Rev.\ D {\bf 93} (2016) no.3,  035012
  doi:10.1103/PhysRevD.93.035012
  [arXiv:1601.01163 [hep-ph]].


\bibitem{Feng:2003xh}
  J.~L.~Feng, A.~Rajaraman and F.~Takayama,
  Phys.\ Rev.\ Lett.\  {\bf 91} (2003) 011302
  doi:10.1103/PhysRevLett.91.011302
  [hep-ph/0302215].
  
\bibitem{Hessler:2014ssa}
  A.~G.~Hessler, A.~Ibarra, E.~Molinaro and S.~Vogl,
  Phys.\ Rev.\ D {\bf 91} (2015) no.11,  115004
  doi:10.1103/PhysRevD.91.115004
  [arXiv:1408.0983 [hep-ph]].
  
\bibitem{Chatrchyan:2013oca}
  S.~Chatrchyan {\it et al.} [CMS Collaboration],
  JHEP {\bf 1307} (2013) 122
  doi:10.1007/JHEP07(2013)122
  [arXiv:1305.0491 [hep-ex]].
  
\bibitem{Aad:2013gva}
  G.~Aad {\it et al.} [ATLAS Collaboration],
  Phys.\ Rev.\ D {\bf 88} (2013) no.11,  112003
  doi:10.1103/PhysRevD.88.112003
  [arXiv:1310.6584 [hep-ex]].
  
\bibitem{Khachatryan:2015lla}
  V.~Khachatryan {\it et al.} [CMS Collaboration],
  Eur.\ Phys.\ J.\ C {\bf 75} (2015) no.7,  325
  doi:10.1140/epjc/s10052-015-3533-3
  [arXiv:1502.02522 [hep-ex]].
  
  


\bibitem{Belyaev:2012qa}
  A.~Belyaev, N.~D.~Christensen and A.~Pukhov,
  Comput.\ Phys.\ Commun.\  {\bf 184} (2013) 1729
  doi:10.1016/j.cpc.2013.01.014
  [arXiv:1207.6082 [hep-ph]].
  

  
\bibitem{Sjostrand:2014zea}
  T.~Sjostrand {\it et al.},
  Comput.\ Phys.\ Commun.\  {\bf 191} (2015) 159
  doi:10.1016/j.cpc.2015.01.024
  [arXiv:1410.3012 [hep-ph]].
  
\bibitem{deFavereau:2013fsa}
  J.~de Favereau {\it et al.} [DELPHES 3 Collaboration],
  JHEP {\bf 1402} (2014) 057
  doi:10.1007/JHEP02(2014)057
  [arXiv:1307.6346 [hep-ex]].
  
  
\bibitem{Feldman:1997qc}
  G.~J.~Feldman and R.~D.~Cousins,
  Phys.\ Rev.\ D {\bf 57} (1998) 3873
  doi:10.1103/PhysRevD.57.3873
  [physics/9711021 [physics.data-an]].
  
  
\bibitem{Chou:2016lxi}
  J.~P.~Chou, D.~Curtin and H.~J.~Lubatti,
  arXiv:1606.06298 [hep-ph].
  
  
\bibitem{Aad:2014vma}
  G.~Aad {\it et al.} [ATLAS Collaboration],
  JHEP {\bf 1405} (2014) 071
  doi:10.1007/JHEP05(2014)071
  [arXiv:1403.5294 [hep-ex]].
  
  
  
  
\bibitem{Drees:2013wra}
  M.~Drees, H.~Dreiner, D.~Schmeier, J.~Tattersall and J.~S.~Kim,
  Comput.\ Phys.\ Commun.\  {\bf 187} (2014) 227
  doi:10.1016/j.cpc.2014.10.018
  [arXiv:1312.2591 [hep-ph]].


\bibitem{Khachatryan:2014mea}
  V.~Khachatryan {\it et al.} [CMS Collaboration],
  Phys.\ Rev.\ Lett.\  {\bf 114} (2015) no.6,  061801
  doi:10.1103/PhysRevLett.114.061801
  [arXiv:1409.4789 [hep-ex]].


\bibitem{CMS:2014hka}
  V.~Khachatryan {\it et al.} [CMS Collaboration],
  Phys.\ Rev.\ D {\bf 91} (2015) no.5,  052012
  doi:10.1103/PhysRevD.91.052012
  [arXiv:1411.6977 [hep-ex]].

\bibitem{Casas:2001sr}
  J.~A.~Casas and A.~Ibarra,
  Nucl.\ Phys.\ B {\bf 618} (2001) 171
  doi:10.1016/S0550-3213(01)00475-8
  [hep-ph/0103065].


  
\bibitem{Agashe:2014kda}
  K.~A.~Olive {\it et al.} [Particle Data Group Collaboration],
  Chin.\ Phys.\ C {\bf 38} (2014) 090001.
  doi:10.1088/1674-1137/38/9/090001



  

\bibitem{Aad:2008zzm}
  G.~Aad {\it et al.} [ATLAS Collaboration],
  JINST {\bf 3} (2008) S08003.
  doi:10.1088/1748-0221/3/08/S08003
  


\bibitem{ATLAS:1996ab}
  [ATLAS Collaboration],
  CERN-LHCC-96-41.


\bibitem{ATLAS:1996aa}
  [ATLAS Collaboration],
  CERN-LHCC-96-42.
  
  
\bibitem{ATLAS:1997ad}
  [ATLAS Collaboration],
  CERN-LHCC-97-22, ATLAS-TDR-10.
  
\bibitem{Groom:2001kq}
  D.~E.~Groom, N.~V.~Mokhov and S.~I.~Striganov,
  Atom.\ Data Nucl.\ Data Tabl.\  {\bf 78} (2001) 183.
  doi:10.1006/adnd.2001.0861

  
  
\bibitem{Aurousseau:2010oaa}
  M.~Aurousseau,
  CERN-THESIS-2010-138, LAPP-T-2010-05.


\bibitem{Abdelalim:2008gsa}
  A.~A.~Abdelalim,
  CERN-THESIS-2008-151.


  
  
\bibitem{Liu:2015bma}
  Z.~Liu and B.~Tweedie,
  JHEP {\bf 1506} (2015) 042
  doi:10.1007/JHEP06(2015)042
  [arXiv:1503.05923 [hep-ph]].

\end{thebibliography}
\end{document}